 \documentclass{wiley-article}
 \usepackage[numbers]{natbib}
\usepackage{siunitx}

\newcommand{\bsb}{\boldsymbol{\beta}}
\newcommand{\bsl}{\boldsymbol{\lambda}}
\newcommand{\bsL}{\boldsymbol{\Lambda}}
\newcommand{\bsP}{\boldsymbol{\Psi}}

\newcommand{\bsS}{\boldsymbol{\Sigma}}

\newcommand{\bsp}{\boldsymbol{\Phi}}
\newcommand{\bsth}{\boldsymbol{\theta}}

\newcommand{\bsd}{\boldsymbol{\delta}}
\newcommand{\bsD}{\boldsymbol{\Delta}}

\newcommand{\R}{\ensuremath{\mathbb{R}}}

\newcommand{\mbf}{\mathbf{f}}
\newcommand{\mbb}{\mathbf{b}}

\newcommand{\mbB}{\mathbf{B}}
\newcommand{\mbx}{\mathbf{x}}
\newcommand{\tx}{\widetilde{x}}
\newcommand{\tmbx}{{\widetilde{\mathbf{x}}_{is}}}

\newcommand{\mbF}{\mathbf{F}}
\newcommand{\mbX}{\mathbf{X}}

\newcommand{\mbI}{\mathbf{I}}

\newcommand{\mbE}{\mathbf{E}}
\newcommand{\mbl}{\mathbf{l}}
\newcommand{\mbL}{\mathbf{L}}
\newcommand{\mbC}{\mathbf{C}}

\newcommand{\mbR}{\mathbf{R}}

\newcommand{\mbW}{\mathbf{W}}

\newcommand{\mbe}{\mathbf{e}}
\newcommand{\Var}{\rm{Var}}
\newcommand{\Cov}{\rm{Cov}}
\newcommand{\Exp}{\rm{E}}

\newcommand{\new}[1]{{\color{black} #1}}
\papertype{Research Article}

\title{Multi-study factor regression model: an application in nutritional epidemiology}

\abbrevs{MSFR, Multy-Study Factor Regression; ECM, Expectation/Conditional Maximization.}

\author[1,2]{Roberta De Vito}
\author[3,4]{Alejandra Avalos-Pacheco}

\affil[1]{Department of Biostatistics, Brown University, Providence, RI, 02903, USA}
\affil[2]{Data Science Institute, Brown University, Providence, RI, 02906, USA}
\affil[3]{Institute of Applied Statistics, Johannes Kepler University Linz, 4040 Linz, Austria}
\affil[4]{Harvard-MIT Center for Regulatory Science, Harvard University, Boston, MA, 02115, USA}

\corraddress{Roberta De Vito PhD, Brown University, Providence, RI, 02906, USA}
\corremail{roberta\_devito@brown.edu}

\fundinginfo{RDV was supported by the US National Institutes of Health, NIGMS/NIH COBRE CBHD P20GM109035}

\runningauthor{De Vito \& Avalos-Pacheco}

\begin{document}

\begin{frontmatter}
\maketitle

\begin{abstract}
Diet is a risk factor for many diseases. In nutritional epidemiology, studying reproducible dietary patterns is critical to reveal important associations with health. However, this task is challenging: diverse cultural and ethnic backgrounds may critically impact eating patterns by showing heterogeneity, leading to incorrect dietary patterns and obscuring the components shared across different groups or populations. Moreover, covariate effects generated from observed variables, such as demographics and other confounders, can further bias these dietary patterns. Identifying the shared and group-specific dietary components and covariate effects is essential to drive accurate conclusions. To address these issues, we introduce a new modeling factor regression, the Multi-Study Factor Regression (MSFR) model. The MSFR model analyzes different populations simultaneously, achieving three goals: capturing shared component(s) across populations, identifying group-specific structures, and correcting for covariate effects. We use this novel method to derive common and ethnic-specific dietary patterns in a multi-center epidemiological study in Hispanic/Latinos community. Our model improves the accuracy of common and group dietary signals, provides a robust estimation of factor cardinality, and yields better prediction than other techniques, revealing important associations with health and cardiovascular disease. In summary, we provide a tool to integrate different groups, providing accurate dietary signals crucial to inform public health policy.

\keywords{Factor analysis, Joint analysis, ECM algorithm, Nutritional epidemiology, Dietary patterns}
\end{abstract}
\end{frontmatter}

\section{Introduction}
\label{sec:intro}

The impact of diet on diseases such as cancer \citep{willett1984diet} and cardiovascular disease (CVD) \citep{kromhout2001diet,gbd2015global,aune2016nut} has continually affected populations worldwide, including the United States. Cardiovascular disease alone accounted for over 900,000 deaths in 2021 \citep{nat_2024}. Therefore, detecting eating patterns is crucial for assessing the association between diet and disease, and guiding and informing public health policy.

Diverse cultural and ethnic background components may have crucial impacts on eating patterns. 
In nutritional epidemiology, heterogeneous high-dimensional data coming from different populations or ethnic backgrounds \citep{stearns2017ethnic, zulyniak2017does} often arise \citep{haines1999diet}, providing an opportunity to estimate shared dietary patterns and differences across them, leading to a better understanding of their association with certain types of diseases. 

When analyzing different populations or groups, some diets and foods are reproducible and shared, while other diet characteristics are specific to each population. 
To handle this task, standard procedures involve stacking all the populations and performing traditional techniques such as factor analysis \citep{haines1999diet,schulze2003approach, edefonti2012nutrient, bennett2022comparison} to estimate dietary patterns, or performing factor analysis in each population and then adopting a measure of correlation to identify the most similar dietary patterns across the different populations or subgroups \citep{castello2014spanish, castello2022adherence}.
However, identifying dietary patterns in different populations can be challenging, mainly due to the diversity of ethnicities and demographic compositions among populations. 
Consumption patterns of foods, recipes, or beverages can vary significantly between populations, critically affecting the risk of disease outcomes. 
Population-based identification of these diets is crucial to provide essential knowledge and information for broad public health policies. 
However, distinctly unique populations can impact and mask the shared and reproducible dietary patterns among all the populations. Therefore, there is a crucial need for rigorous statistical techniques to identify concurrently common and study-specific diet characteristics to clearly estimate the association between diet and diseases.

Researchers in the field have been increasingly focused on identifying both group-specific and reproducible signal patterns. The multi-study factor analysis (MSFA) model \citep{DeVito2019Mfa} introduced a framework for systematically separating shared and study-specific characteristics in multi-study settings, such as gene expression data. 

Building on this, subsequent methods address the identification of common and group-specific signals. For example, \cite{roy2021perturbed} proposed a technique that perturbs the data by multiplying it with a group-specific matrix to identify a common structure. However, this approach estimates only common factors adjusted for study-specific perturbations. More recently, the SUbspace Factor Analysis (SUFA) model \citep{chandra2024inferring} extends this direction by estimating both common and study-specific factors. SUFA identifies a common low-dimensional subspace while simultaneously capturing group-level variation at the subspace level.

The MSFA approach was also adopted and extended in nutritional epidemiology. 
For instance, \cite{de2019shared} analyzed the International Head and Neck Cancer Epidemiology (INHANCE) consortium \citep{hashibe2007alcohol, conway2009enhancing} data via MSFA, revealing important common and population-specific diet characteristics and their association with the two cancer types.
In this direction, \cite{de2022shared} adopted a multi-study Bayesian approach in a nutritional epidemiological setting in 4 US field sites (Bronx, Chicago, Miami, San Diego)  in the Hispanic Community Health Study/Study of Latinos (HCHS/SOL) \citep{sorlie2010design}.  
More recently, \cite{stephenson2020robust} introduced a generalization of mixture models, namely the Robust Profile Clustering (RPC), to handle diverse populations or studies in a case-control birth defects study in the United States \citep{yoon2001national}. 
The RPC identifies robust food ``global'' clusters across all populations and local population clusters.
A recent extension of RPC incorporates a supervised learning approach \citep{stephenson2022derivation}, adjusting for the association between diet and orofacial cleft birth defects while accounting for additional confounders.

All of these approaches highlight the crucial need in nutritional epidemiology to identify common and study diet components concurrently, in order to gain a better understanding of their association  with disease outcomes.
However, none of these approaches accounts for covariates or confounders essential to identify the diet signals accurately.  Specifically, the diet is affected by numerous confounders \citep{trichopoulos1985diet, tucker1997body}, such as smoking, alcohol, gender, and other covariates, which can obscure both the common and study-specific signals. Thus, not adjusting for multiple potential confounding factors could result in inaccurate conclusions.

In the standard factor analysis framework, models that account for observed covariates are well-studied. 
For instance, \cite{2003Bfrm} introduced sparse latent factor models with many explanatory variables. 
This approach enables a more precise assessment of the factor structure, leading to improved breast cancer data predictions.
\cite{Carvalho2008HSFM} developed a factor regression model that estimated both factor loadings and regression coefficients. 
This paper provided a modeling approach that improved the characterization of the factor loading patterns in cancer genomic applications by accounting for observed covariates. 
This approach also showed an increased predictive ability when considering observed covariates. 
More recently, \cite{Avalos2022HLDI} 
introduced the Bayesian factor regression model, a novel approach that jointly analyzes multiple studies and adjusts for systematic biases, such as batch effects, in a latent factor regression framework. The authors illustrated their analysis' power in both signal detection and predictive ability and empirically showed that inaccurate conclusions can be drawn if batch effects are not appropriately considered. 

Although all these methods revealed the crucial need to account for both latent structures and observed variables, none of them considered study-specific latent patterns that contribute significantly to the variations of each population. 
Indeed, these populations or ethnic background dietary patterns can reveal essential differences among groups and sometimes mask the shared signal among groups, drastically biasing the associations with the disease.
Therefore, it is critical to identify both the shared and study-specific latent components while accounting for other observed covariates, like counfonders and/or batch effects.
To address this challenge, we propose a new modeling approach, the Multi-Study Factor Regression (MSFR) model, which is able to analyze multiple studies/ populations/groups jointly and estimate three different components: (1) the reproducible and shared latent components, (2) the study-specific latent component for each population, (3) the regression coefficients, which are key to detect confounders.

Our work focuses on the Hispanic Community Health Study/Study of Latinos (HCHS/SOL) \citep{national2009hispanic}, which is a multi-center epidemiologic study in Hispanic/Latino populations.

The HCHS/SOL aims to detect crucial factors impacting the health of Hispanics/Latinos \citep{sorlie2010design}. One of the primary objectives of the study is to explore how dietary patterns influence cardiovascular disease risk \citep{daviglus2014cardiovascular} across various Hispanic/Latino groups, including Cuban, Puerto Rican, Dominican, Mexican, and Central/South American populations. In this paper, we will refer to these Hispanic/Latino groups collectively as ethnic backgrounds.

Our MSFR model aims to identify the reproducible dietary components among the different ethnic backgrounds and background dietary components while accounting for confounders (i.e., smoking, alcohol, etc.) that can obscure the two dietary signals. In this way, we can assess a more accurate association between dietary patterns and cardiovascular diseases. 

The plan of the paper is as follows. Section~2 motivates our new approach by describing the Hispanic Community Health Study/Study of Latinos (HCHS/SOL) and all the variables considered in our analysis. Section~3 presents the multi-study regression approach, including the model definition and model estimation. Section~4 presents the performance of our MSFR in different simulation settings, comparing it with the multi-study factor model, the factor regression model and a two-step approach aimed to correct for covariate effects. Section~5 includes the application of MSFR in the HCHS/SOL study, which motivated our novel method. Section~6 includes the discussion.

\section{The Hispanic Community Health Study/Study of Latinos (HCHS/SOL)}

\subsection{Design of the study}\label{sec2}
 The information for this study was collected at baseline (2008–2011) using a stratified two-stage probability sampling design, as described in detail by LaVange et al. (2010). Data collection involved two 24-hour dietary recalls: one conducted in person at baseline and the other conducted via telephone within 30 days of the baseline assessment. Interviews were conducted in the participant’s preferred language, with 20\% of participants opting for English and 80\% for Spanish, ensuring cultural and linguistic appropriateness.

The dietary data, derived from thousands of common and Hispanic/Latino foods, includes 139 nutrients,  132 food-group serving counts, nutrient ratios, and other food components. Almost all participants (99\%) provided at least one dietary recall, ensuring a comprehensive dataset.

\subsection{Data preprocessing}\label{sec2}

A total of 14,002 participants' data were collected. We excluded participants who self-identified as belonging to other ethnic backgrounds and/or present recall data unreliable (i.e., vitamin E $< 0$) and/or show inconsistent (i.e., $< 1st \mbox{ or } > 99th $percentiles) estimates of energy intake. We also removed individuals with missing ethnic background information. After applying all of the above-mentioned exclusion criteria, we performed our analyses on a total of 10,460 participants.

We formed 25 food groups that best represent the overall diet for Hispanics/Latinos based on nutritional, cultural, and behavioral importance \citep{sofianou2011differences, batis2011food, davis2013dietary} (details in Table~S1, Supp Materials \S~A). Specifically, we aggregated ingredients (i.e., tortilla, beef, processed meat) composing a recipe (i.e., taco) and nonrecipe foods (i.e., Dark Green Veg, tomato, etc.) to reflect eating attitudes \citep{maldonado2022posteriori}. 
This is the first time that food groups have been analyzed in a multi-study setting while correcting for covariate effects.  \citet{maldonado2022posteriori} estimates different patterns from food groups to understand differences across ethnic backgrounds by using a standard FA in each of the ethnic background groups. However, in this analysis, shared components are not estimated and established: these components are crucial to fully understanding the cultural and ethnic background diet.

We derived food groups from the mean of the two available, reliable recalls collected at baseline (2008-2011) to  include all dietary information. We log-transformed the food group intakes and standardized the data.

An overview of socio-demographic characteristics for each ethnic background group is provided in  Table~\ref{tab:SOL}. Note that we only report socio-demographic characteristics included in our model as counfonders (Section~\ref{sub:HCHS/SOL}).

\begin{table*}[h]
  \caption{\it Social and Demographic characteristic of the HCHS/SOL consortium for each ethnic background. }\label{tab:SOL}
    \tabcolsep=0pt
    \begin{tabular*}{\textwidth}{@{\extracolsep{\fill}}c c  c  c  c  c  c@{\extracolsep{\fill}}}
    \hline \hline
     &  Dominica  & Central America & Cuba & Mexico & Puerto Rico  & South America \\
    &   ($n=1005$)  & ($n=1788$) & ($n=1555$) & ($n=3998$) & ($n=1806$) & ($n=308$)\\
      \hline \hline 
      \textbf{BMI$^{1}$}\\mean(sd)
       & 29.4(5.5) & 29.1(5.4)& 29.2(5.5) & 29.5(5.5) & 30.8(6.7) & 30.0 (5.9)\\
      \hline 
     \textbf{Gender} &  &  &  &  & &\\
     Female & 638 & 1064  & 818 & 2441  & 1062& 170\\
     Male & 367 & 724 & 737 & 1557 & 744 & 138\\
     \hline
      \textbf{Employment} &  &  &  &  & &\\
      Retired & 113 & 116 & 159 & 271 & 405 & 22\\
      Not retired  and
      not employed & 386 & 606 & 717 & 1411 & 722 & 123\\
      Part-time ($\leq$ 35 hr) & 146 & 394 & 191 & 807 & 164 & 51\\
      Full-time ($>$ 35 hr) & 360 & 672 & 488 & 1509 & 515 & 112\\
\hline 
      \textbf{Education} &  &  &  &  & &\\
      $<$ High School & 437 & 615 & 354 & 1768 & 694 & 80\\
      = High School & 197 & 427 & 441 & 1029 & 473 & 59\\
      $>$ High School & 371 & 746 & 760 & 1201 & 639 & 196\\
      \hline
      \textbf{Yearly Household Income}  &  &  &  &  & &\\
      $<$ 30k & 706 & 1207 & 1058 & 2442 & 1121 & 165\\
      $\geq$ 30k & 241  & 469  & 352 & 1415  & 602 & 124\\
      Not reported & 58 & 112 & 145 & 141 & 83 & 19\\
        \hline
       \textbf{Antihypertensives} &  &  &  &  & &\\
     No & 805 & 1553  & 1207 & 3418  & 1376 & 276\\
     Yes & 200 & 235 & 348 & 580 & 430 & 32\\
      \hline
       \textbf{Antidiabetics} &  &  &  &  & &\\
     No & 892 & 1638  & 1404 & 3513  & 1491 & 282\\
     Yes & 113 & 150 & 151 & 485 & 315 & 26\\
     \hline 
      \textbf{Years lived in USA} &  &  &  &  & &\\
      $<$ 10 years & 235 & 573 & 744 & 764 & 85 & 40\\
      $\geq$ 10 and $<$ 20 years & 296 & 541 & 412 & 904 & 107 & 47\\
      $\geq$ 20 years & 398 & 598 & 339 & 1699 & 874 & 59\\
      US Born & 76 & 76 & 60 & 631 & 740 & 162\\
      \hline 
      \textbf{Marital Status} &  &  &  &  & &\\
      Single & 338 & 463 & 323 & 750 & 686 & 124\\
      Married & 430 & 959 & 849 & 2537 & 675 & 127\\
      Separated & 237 & 366 & 383 & 711 & 445 & 57\\
      \hline 
      \textbf{Physical Activity$^{2}$} &  &  &  &  & &\\
      No & 481 & 1104 & 1229 & 2447 & 1013 & 152\\
      Yes & 524 & 684 & 326 & 1551 & 793 & 156\\
      \hline 
      \textbf{Age}\\mean(sd)
       & 46.2(14.2) & 46.3(13.1)& 50.3(12.6) & 45.7(13.4) & 49.0(14.0) & 40.5 (15.4)\\
      \hline
      \hline
    \end{tabular*}
    \begin{tablenotes}%
\item[$^{1}$] Body Mass Index (kg/$m^2$)
\item[$^{2}$] Met 2008 Physical Activity Guidelines for Americans 
\vspace*{6pt}
\end{tablenotes}
\end{table*}

Ethnic backgrounds were not equally distributed at each study site. Specifically, the population with Mexican (South American) background is the largest (smaller) among the groups. The majority of individuals were female taking no medication for hypertension and diabetes.
The BMI is higher for Puerto Rican compared to other ethnic backgrounds. 
The category of Education most concentrated was ``less than high school'', except for the Central American, Cuban and South American backgrounds with a concentration on ``greater than high school'', similarly for the employment category.  The nativity (i.e., years lived in USA) is very different across the ethnic backgrounds: the majority of individuals from Dominican, Central American, Mexico and Puerto Rican live in USA from more than 20 years, while the majority of individuals from Cuba live in USA from less than 10 years, the majority of individuals from South American are US born.
The covariates further corroborate the heterogeneity across ethnic background. 

We consider a total of 25 food group variables and 11 confounders to best identify the common signals across the 6 different ethnic background groups and the study-specific components for each of these groups. 

\subsection{Cardiovascular disease outcomes}\label{secCVD}
We analyze three baseline outcomes (2007–2011) associated with dietary patterns derived from food groups: diabetes, hypercholesterolemia, and hypertension.

Diabetes is a categorical variable with three levels: (1) normal glucose regulation, (2) impaired glucose tolerance, and (3) diabetes, reflecting progressive stages of the condition. This classification incorporates serum glucose levels adjusted for fasting time, post-OGTT glucose levels (when available), and records of anti-diabetic medication use. Participants without glucose laboratory data and no recorded use of anti-diabetic medications were assumed to have normal glucose regulation. For analysis, diabetes was dichotomized: levels 1 and 2 were combined as 0 (no diabetes), and level 3 was classified as 1 (diabetes). Both type 1 and type 2 diabetes are strongly associated with cardiovascular disease (CVD) \citep{matheus2013impact}.

Hypercholesterol is a binary variable determined by total cholesterol, HDL, LDL levels, and the use of antihyperlipidemic medications. High cholesterol levels linked to dietary habits significantly increase CVD risk \citep{berger2015dietary}.

Hypertension is defined as a binary variable: systolic or diastolic blood pressure $\geq$ 140/90 mmHg or the use of antihypertensive medications (scanned records only). Hypertension is a key contributor to CVD risk \citep{sowers2001diabetes}.

These variables are essential for evaluating whether dietary patterns, including those specific to ethnic backgrounds, are associated with cardiometabolic disease risk \citep{liese2022variations}.

\section{Multi-study factor regression model}

We jointly model the data from $S$ different groups, populations, or studies that observe the same $p$ (i.e., food). We consider vectors $\mbx_{is} \in \R^p$, observed for $i = 1, \dots, n_s$ individuals in $s=1,\dots,S$ studies. 
Our method generalizes the traditional multi-study factor analysis (MSFA) \citep{DeVito2021Bmfa}.

Traditional MSFA models $\mbx_{is}$ as a linear combination of low-dimensional common factors $\mbf_{is} \in \R^q$, and group-specific factors $\mbl_{is} \in \R^{q_s}$, $q + q_s \ll p$:
\begin{align}
\label{eq:MSFA}
    \mbx_{is} = \bsp \mbf_{is} + \bsL_s \mbl_{is} + \mbe_{is},
\end{align}
where $\bsL_s \in \R^{p \times p_s}$ for $s=1,\dots,S$, is the study-specific loadings matrix, $\bsp \in \R^{p \times q}$ is the factor loadings matrix, and $\mbe_{is} \in \R^p$ is the error, distributed as $\mbe_{is} \sim N(0,\bsP_s), s=1,\dots,S$ where $\bsP_s = \text{diag}\{ \psi_{js} \}_{j=1}^p$ is the idiosyncratic error covariance matrix of study $s$. 
Common and study-specific factors are assumed to be standard normal and independent between them and the errors. 
Marginally, observations $\mbx_{is}$ are normally distributed, centered at zero and with covariance matrix $\bsS_s = \bsp \bsp^\top + \bsL_s \bsL_s^\top + \bsP_s$. 
The covariance matrix allows decomposing the total variance into three distinct parts, i.e.\ the variance of the common factors $\bsp \bsp^\top$, of the study-specific factors $\bsL_s \bsL_s^\top$ and the error $\bsP_s$.

Model~\eqref{eq:MSFA} assumes that the data are explained by unobserved latent factors only, without taking into account prespecified variables that can affect the outcome in practice. 
For instance, in our HCHS/SOL application, the MSFA ignores the effect that observed covariates, such as BMI, alcohol and cigarette use, can have on cardiovascular-disease-related dietary habits. 
In other words, MSFA models the data without allowing for other additional regression effects that do not affect the underlying dietary patterns given in the common and study-specific factors. 
We will later describe how much inference can suffer when not incorporating such covariate adjustments (Table~\ref{tab:MSE}, Figures \ref{fig:BoxCardinality}, \ref{fig:HeatAll},  and \ref{fig:BoxPlotsAll}, 
and Supplementary Materials). 
We tackle this issue by extending Model~\eqref{eq:MSFA} to incorporate covariate adjustments by adopting a factor regression approach \citep{Avalos2022HLDI, Avalos2018Frfd, Carvalho2008HSFM}. 
Let $\mbb_{is}$ be $p_b$-dimensional vector of known covariates. 
The Multy-study Factor Regression model is:
\begin{equation}
\label{eq:model}
\mbx_{is} = \bsb \mbb_{is} + \bsp \mbf_{is} + \bsL_s \mbl_{is} + \mbe_{is}
\end{equation}
where 
$\bsb \in \R^{p \times p_b}$ is the matrix of regression coefficients. 

We remark that the utility of this model goes beyond our Hispanic Community Health Study. 
For instance, this model can easily adjust for several covariate effects, such as complex batch effects, by simply setting $\mbb_{is} \in [0,1]^{S}$ as a study indicator. 
In this setting, $\beta$ captures the mean batch effect adjustment and $\bsP_s$ the variance adjustment.
Furthermore, each column of \(\beta\) can represent the covariate effect specific to a study, as demonstrated in Scenario 3 of the simulation studies. To achieve this, we allow \(b_{is}\) to be a vector of length \(p_b=S\), where \(b_{is} \neq 0\) if individual $i$ belongs to study $s$, and \(b_{is} := 0\) otherwise. This framework can also be extended by partitioning the covariate effects \(p_b = p_{b_1} + p_{b_2} + \dots + p_{b_S}\)  allowing study-specific non-zero values while keeping the rest at zero.

\section*{Maximum likelihood estimation by ECM algorithm}
We fit Model~\eqref{eq:model} using a deterministic Expectation/Conditional Maximization (ECM) optimization \citep{MengECM93}, along the lines of \cite{DeVito2019Mfa} to estimate the parameters $\bsth= \left\{\bsp, \bsL_s, \bsP_s, \new{\bsb} \right\}$. 
Our proposed ECM, summarized in Algorithm~\ref{alg:ECM},  maximises the
expected complete-data log-likelihood $l_c(\bsth)=E_{\mbF,\mbL_s\mid \bsth,\mbX}[p(\bsth \mid \mbF,\mbL_s,\mbX)]$  iteratively. 
For simplicity, let $\tmbx = \mbx_{is}\new{- \bsb \mbb_{is} }$, and $\widehat{\bsth}=\left\{\widehat{\bsp}, \widehat{\bsL}_s, \widehat{\bsP}_s,\widehat{\bsb} \right\}$ be the current value of the estimated parameters.

\begin{algorithm*}[!t]
  \textbf{initialize} $\widehat{\bsp}=\bsp^{(0)}, \widehat{\bsL}_s=\bsL_s^{(0)}, \widehat{\bsP}_s=\bsP_s^{(0)}, \widehat{\bsb}=\bsb^{(0)}$\\
  
 \textbf{while }{$t<T$ and $\epsilon>\epsilon^*$}

\begin{mdframed}[linewidth=1,linecolor=black, innertopmargin=-.1em, innerbottommargin=0em, topline=false, rightline=false,bottomline=false]
  
\textbf{E-step}$^\S$:\\
$$\begin{array}{cccc}
& E[\mbf_{is} |\tmbx, \widehat{\bsth}] = \mbE_{f_{i}} =  \bsd \tmbx, & & E[\mbl_{is} |\tmbx, \widehat{\bsth}] =\mbE_{l_{is}} = \bsd_s \tmbx, \\
& \Exp\left[\mbC_{\tx_s\tx_s} | \tmbx, \widehat{\bsth} \right] = \mbE_{\tx_s\tx_s} = \mbC_{\tx_s\tx_s},  & & \Exp\left[\mbC_{\tx_sl_s} |\tmbx, \widehat{\bsth} \right] =  \mbE_{\tx_sl_s} = \mbC_{\tx_s\tx_s}\bsd_{s}^\top,\\
&\Exp\left[\mbC_{\tx_sf} | \tmbx, \widehat{\bsth}\right] = \mbE_{\tx_sf} = \mbC_{\tx_s\tx_s}\bsd^\top, & & \Exp\left[\mbC_{ff} | \tmbx, \widehat{\bsth} \right]= \mbE_{ff} = \bsd \mbC_{\tx_s\tx_s}\bsd^\top + \bsD,\\
& \Exp\left[\mbC_{l_sl_s} | \tmbx, \widehat{\bsth} \right] =\mbE_{l_sl_s} = \bsd_{s} \mbC_{\tx_s\tx_s}\bsd_{s}^\top + \bsD_{s} & & \Exp \left[\mbC_{fl_s} | \tmbx, \widehat{\bsth} \right] = \mbE_{fl_s} = \bsd \mbC_{\tx_s\tx_s}\bsd_{s}^\top + \bsD_{fl}.
\end{array}
$$

 \textbf{CM-step}:\\
  \begin{tabular}{rl}
$\widehat{\bsP}_s=$& 
diag$\left\{ \mbE_{\tx_s\tx_s}  + \widehat{\bsp} \mbE_{f f}  \widehat{\bsp}^{\top} + \widehat{\bsL}_s   \mbE_{l_s l_s}     \widehat{\bsL}_s^{\top}  -  2 \mbE_{\tx_sf } \widehat{\bsp}^{\top} -\Biggl. 2\mbE_{\tx_s l_s}\widehat{\bsL}_s^{\top}  +2 \widehat{\bsp} \mbE_{f l_s}\widehat{\bsL}_s^{\top}\right\}$\\
$\mbox{vec}(\widehat{\bsp}) =$&  $\sum_{s=1}^S \left(\mbE_{ff}^\top \otimes n_s \widehat{\bsP}^{-1}_s\right) 
 \mbox{vec} \left( n_s \widehat{\bsP}^{-1}_s\mbE_{\tx_s f}-n_s\widehat{\bsP}^{-1}_s \widehat{\bsL}_s \mbE_{fl_s}^\top\right)$\\
 $\widehat{\bsL}_s =$&  $  \Biggl( \mbE_{x_sf} - \widehat{\bsp} \mbE_{f l_s}   \Biggr) \Biggl(
  \mbE_{l_s l_s} \Biggr)^{-1}$\\
  $\widehat{\bsb} =$&  $ \left[ \sum_{s=1}^S  \sum_i^{n_s}   (\mbx_{is} - \widehat{\bsp} \mbE_{f_{i}}  - \widehat{\bsL}_{s} \mbE_{l_{is})}  \mbb_{is}^\top \right] \left[ \sum_{s=1}^S  \sum_i^{n_s}   ( \mbb_{is}\mbb_{is}^\top )\right]^{-1}$\\
\end{tabular}

\vspace{.2cm}
\textbf{set}  $\bsth^{(t+1)}=\widehat{\bsth}$\\

\textbf{compute} $t=t+1$  and $\epsilon = \mid l^{(t+1)}-l^{(t)}\mid$, with $l^{(t+1)} = l_c (\bsth^{(t)}) + \frac{l_c (\bsth^{(t+1)}) - l_c (\bsth^{(t)})}{1-c^{(t)}}$, and $c^{(t)}=\frac{l_c (\bsth^{(t+1)}) - l_c (\bsth^{(t)})}{l_c (\bsth^{(t)}) - l_c (\bsth^{(t-1)})}$ 
\end{mdframed}

\textbf{end while}
 
 $^\S$ where
 $$
\begin{array}{cccccc}
& \mbC_{\tx_s\tx_s}  =  \dfrac{\sum_{i=1}^{n_s}\tmbx \tmbx^\top}{n_s}, & & \mbC_{ff} =  \dfrac{\sum_{i=1}^{n_s}\mbf_{is} \mbf_{is}^\top}{n_s}, & & \mbC_{l_sl_s} =  \dfrac{\sum_{i=1}^{n_s}\mbl_{is} \mbl_{is}^\top}{n_s},\\
& \mbC_{\tx_sf}  =  \dfrac{\sum_{i=1}^{n_s}\tmbx \mbf_{is}^\top}{n_s}, & & \mbC_{\tx_sl_s}  =  \dfrac{\sum_{i=1}^{n_s}\tmbx \mbl_{is}^\top}{n_s}, & & \mbC_{fl_s} =  \dfrac{\sum_{i=1}^{n_s}\mbf_{is} \mbl_{is}^\top}{n_s},
\end{array}
$$

$$
\bsD_{fl} = \Cov[\mbl_{is},\mbf_{is}\vert\tmbx, \bsth_{t-1}] = \bsp^\top \bsS_s^{-1} \bsL_s,
$$
 and
$$
\begin{array}{cccc}
& \bsd = \bsp^\top \bsS_s^{-1}, & & \bsD = \Var[\mbf_{is}\vert \tmbx, \bsth_{t-1} ] =  \mathbf{I}_{q}-\bsp^\top \bsS_s^{-1} \bsp, \\
& \bsd_{s} = \bsL_s^\top \bsS_s^{-1},  & & \bsD_{s} = \Var[\mbl_{is}\vert \tmbx, \bsth_{t-1} ] =  \mathbf{I}_{q_s}-\bsL_s^\top \bsS_s^{-1} \bsL_s.
\end{array}
$$
\caption{ECM algorithm for Multi-study Factor Regression}
 \label{alg:ECM}
\end{algorithm*}

In the E-step, we obtained the expected complete-data log-likelihood
\begin{equation}
\begin{split}
\label{eq:ECMcompleteloglike}
l_c (\bsth) =& \sum_{s=1}^S  \Biggl\{    -\frac{n_s}{2} \log \vert \bsP_s \vert  - \frac{n_s}{2} tr \left[\bsP_s^{-1} \frac{\sum_i^{n_s}  \left( E[\tmbx \tmbx^\top |\tmbx, \widehat{\bsth}]    + \bsp E[\mbf_{is}\mbf_{is}^\top |\tmbx, \widehat{\bsth}]\bsp^\top + \bsL_s E[\mbl_{is}\mbl_{is}^\top |\tmbx, \widehat{\bsth}]\bsL_s^\top \right.}  {n_s}\right. \Biggr.\\
&\frac{- 2 E[\tmbx\mbf_{is}^\top |\tmbx, \widehat{\bsth}]\bsp^\top - 2E[\tmbx \mbl_{is}^\top |\tmbx, \widehat{\bsth} ]\bsL_s^\top)  +2  \bsp E[ \mbf_{is} \mbl_{is}^\top |\tmbx, \widehat{\bsth}] \bsL_s^\top )}{n_s} \Biggr]  \Biggr\}+C,
\end{split}
\end{equation}

where $C$ is a constant. 
Equation~\eqref{eq:ECMcompleteloglike} depends on the latent factors $\mbf_{is}$ and $\mbl_{is}$ in terms of their conditional mean and their conditional second moments, shown in Algorithm~\ref{alg:ECM} and Supplement~A. 
All the moments are computed based on the multivariate normal distribution.

Note that the conditional expectations require to invert the $p \times p$ covariate matrix $\widehat{\bsS}_s$, through
\begin{equation}
\begin{array}{cccc}
& \bsd = \widehat{\bsp}^\top \widehat{\bsS}_s^{-1}, & & \bsd_{s} =  \widehat{\bsL}_s^\top \widehat{\bsS}_s^{-1},
\end{array}
\end{equation}
which can be cumbersome and challenging as the number of the variables $p$ increases. 
To address this issue, we apply the Woodbury matrix identity \citep{Morgenstern1950SoIo} twice, obtaining
\begin{equation}
\label{eq:Woodbury}
\begin{split}
 \bsd &= ( \mbI_q +  \widehat{\bsp}^\top  \widehat{\mbW}_{\bsL_s}  \widehat{\bsp})^{-1} \widehat{\bsp}^\top \widehat{\mbW}_{\bsL_s}, \\
 \bsd_{s} &=  ( \mbI_{q_s} +  \widehat{\bsL}_s^\top \widehat{\mbW}_{\bsp}  \widehat{\bsL}_s)^{-1} \widehat{\bsL}_s^\top \widehat{\mbW}_{\bsp},
\end{split}
\end{equation}
with
\begin{equation}
\begin{split}
\label{eq:Woodbury2}
\widehat{\mbW}_{\bsL_s}=&(\widehat{\bsL}_s \widehat{\bsL}_s^\top + \widehat{\bsP}_s)^{-1}
=\widehat{\bsP}_s^{-1} -\widehat{\bsP}_s^{-1} \widehat{\bsL}_s ( \mbI_{q_s} +  \widehat{\bsL}_s^\top \widehat{\bsP}_s^{-1}\widehat{\bsL}_s)^{-1} \widehat{\bsL}_s ^\top \widehat{\bsP}_s^{-1}\\
\widehat{\mbW}_{\bsp} =& (\widehat{\bsp} \widehat{\bsp}^\top + \widehat{\bsP}_s)^{-1}
=\widehat{\bsP}_s^{-1} -\widehat{\bsP}_s^{-1} \widehat{\bsp} ( \mbI_{q} +  \widehat{\bsp}^\top \widehat{\bsP}_s^{-1} \widehat{\bsp})^{-1} \widehat{\bsp} ^\top \widehat{\bsP}_s^{-1}.
\end{split}
\end{equation}
We have carried the inference of $\mbE_{f_{is}}$ and $\mbE_{l_{is}}$ using Equation~\eqref{eq:Woodbury2}. 

In the CM-step, we maximize $l_c (\bsth)$ 
with respect to $\bsp, \bsL_s, \bsP_s, \new{\bsb}$. 
For simplicity, we can re-write Equation~\eqref{eq:ECMcompleteloglike} as
\begin{equation}
\label{eq:ECMcompleteloglike2}
l_c (\bsth) =  \sum_{s=1}^S \left(-\frac{n_s}{2} \log \vert\bsP_s  \vert  - \frac{n_s}{2} tr \left[\bsP_s ^{-1} \left( \mbE_{\tx_s\tx_s} + \bsp \mbE_{f_{s} f_{s}}\bsp^t   + \bsL_s \mbE_{l_{s}l_{s}}\bsL_s^\top -  2 \mbE_{\tx_{s}f_{s}}\bsp^\top - 2\mbE_{\tx_{s} l_{s}}\bsL_s^\top + 2  \bsp \mbE_{f_{s} l_{s}} \bsL_s^\top \right) \right]   \right),
\end{equation}
with $\mbE_{\tx_s\tx_s}, \mbE_{f_{s} f_{s}}, \mbE_{l_{s}l_{s}}, \mbE_{\tx_{s}f_{s}}$ and $\mbE_{f_{s} l_{s}}$ as in Algorithm~\ref{alg:ECM}.
We obtain the updates for each parameter setting the partial derivatives of $l_c (\bsth)$ to zero, and update each of them iteratively while keeping the others fixed at their most recent update. 
The updates for the precision $\bsP_s$, common $\bsp$, and study-specific loadings matrices $\bsL_s$ are obtained analogously to the ECM updates in the MSFA  \citep{DeVito2019Mfa}, with $\tmbx = \mbx_{is}\new{- \bsb \mbb_{is} }$, maximizing Equation~\eqref{eq:ECMcompleteloglike2} w.r.t. $\bsp, \bsL_s, \bsP_s$:
\begin{equation}
    \begin{split}
     \widehat{\bsP}_s =&  
\text{diag}\left\{ \mbE_{\tx_s\tx_s}  + \widehat{\bsp} \mbE_{f f}  \widehat{\bsp}^{\top} + \widehat{\bsL}_s   \mbE_{l_s l_s}     \widehat{\bsL}_s^{\top}   -  2 \mbE_{\tx_sf } \widehat{\bsp}^{\top} -\Biggl. 2\mbE_{\tx_s l_s}\widehat{\bsL}_s^{\top}  +2 \widehat{\bsp} \mbE_{f l_s}\widehat{\bsL}_s^{\top}\right\}\\
\mbox{vec}(\widehat{\bsp}) =&  \sum_{s=1}^S \left(\mbE_{ff}^\top \otimes n_s \widehat{\bsP}^{-1}_s\right) 
 \mbox{vec} \left( n_s \widehat{\bsP}^{-1}_s\mbE_{\tx_s f}- n_s\widehat{\bsP}^{-1}_s \widehat{\bsL}_s \mbE_{fl_s}^\top\right)\\
 \widehat{\bsL}_s =&    \Biggl( \mbE_{x_sf} - \widehat{\bsp} \mbE_{f l_s}   \Biggr) \Biggl(
  \mbE_{l_s l_s} \Biggr)^{-1},
    \end{split}
\end{equation}

where $\otimes$ is the Kronecker product, vec is the vec operator, and the linear equation is solved with the Lyapunov Equation. More details about the optimizations can be found in Supplementary A.

To estimate the coefficients $\bsb$, we first expand Equation~\eqref{eq:ECMcompleteloglike}:
\begin{equation}
\label{eq:completeloglikebeta0}
l_c (\bsth)
\propto  \sum_{s=1}^S  \left\{ - \frac{n_s}{2} tr \left[\bsP_s^{-1} \frac{\sum_i^{n_s}  \left( -2 \mbx_{is} \mbb^\top_{is}\bsb^\top + \bsb \mbb_{is}\mbb_{is}^\top\bsb^\top +2 \bsb \mbb_{is}\mbE_{f_{i}}^\top \bsp^\top + 2\bsb \mbb_{is} \mbE_{l_{is}}^\top\bsL_s^\top\right)}{n_s} \right]\right\}.\\
\end{equation}

We then take the derivative and the expected values of Equation~\eqref{eq:completeloglikebeta0}
\begin{equation}
\label{eq:derivBetaO}
\frac{\partial l_c (\bsth)}{\partial \bsb} = \sum_{s=1}^S  \left( - \frac{n_s}{2} \bsP_s^{-1}  \left[\frac{\sum_i^{n_s}  2 \left( \bsb \mbb_{is} \mbb_{is}^\top - (\mbx_{is} - \bsp \mbE_{f_{i}} - \bsL_s \mbE_{l_{is}} \right) \mbb_{is}^\top}{n_s} \right]  \right) = 0,
\end{equation}
and solve Equation~\eqref{eq:derivBetaO} to obtain:
\begin{equation}
\begin{split}
\label{eq:MLEBeta}
\widehat{\bsb} =& \left[ \sum_{s=1}^S  \sum_i^{n_s}   (\mbx_{is} - \widehat{\bsp}\mbE_{f_{i}}  - \widehat{\bsL}_{s} \mbE_{l_{is}})  \mbb_{is}^\top \right] \left[ \sum_{s=1}^S  \sum_i^{n_s}   ( \mbb_{is}\mbb_{is}^\top )\right]^{-1}.
\end{split}
\end{equation}

The stopping criterion used for our proposed  ECM algorithm is based on Aitken acceleration \citep{McLachlan2008TEaa}. 
Algorithm~\ref{alg:ECM} is stopped when a tolerance $\epsilon^* = \mid l^{(t+1)}-l^{(t)}\mid$, with $l^{(t+1)} = l_c (\bsth^{(t)}) + \frac{l_c (\bsth^{(t+1)}) - l_c (\bsth^{(t)})}{1-c^{(t)}}$, and $c^{(t)}=\frac{l_c (\bsth^{(t+1)}) - l_c (\bsth^{(t)})}{l_c (\bsth^{(t)}) - l_c (\bsth^{(t-1)})}$ in the complete log-likelihood change is reached, or a maximum number of iterations $T$ have been completed. 
By default we set $\epsilon^*=10^{-7}$ and $T = 50,000$.

The selection of the dimension of the common and study-specific factors $q$ and $q_s$ is made by choosing the model that leads to the best Akaike information criterion (AIC) \citep{AkaikeH1974Anla} or the Bayesian information criterion (BIC) \citep{SCHWARZG1978EtDo}. More details about the latent cardinality selection will be discussed in Sections \S\ref{sub:Simulations} and \S\ref{sub:HCHS/SOL}.

\subsection{Initialization of the ECM algorithm}
Parameter initialization is crucial to obtain better local modes and reduce computational time, as the ECM algorithm can be sensitive to parameter initialization. 
Here, we propose a simple two-step least-squares initialisation by stacking all the $S$ study-data in a single data set $\mbX^\top \in \R^{n \times p}, n=n_1+\dots+n_S$:
\begin{itemize}
\item \textbf{Step 1}
initialize $\bsb^{(0)}=\left[\mbB \mbB^\top \right]^{-1}\mbB  \mbX^\top$ with $\mbX^\top \in \R^{n \times p}, \mbB^\top \in \R^{n \times p_b}$
\item \textbf{Step 2} let $\widetilde{\mbX} = \mbX - \bsb^{(0)} \mbB$. 
Initialize the extra parameters as the Multi-study factor analysis:
\begin{enumerate}
    \item perform a Principal Components Analysis (PCA) on $\widetilde{\mbX}$ and initialized $\bsp^{(0)}$ with the first $q$ principal components, and
    \item perform factor analysis for each $\widetilde{\mbX}_s, s=1,\dots,S$, and initialize $\bsL^{(0)}$ and $\bsP^{(0)}$ with the obtained loading matrices and unharnesses, respectively.
\end{enumerate}
\end{itemize}

\subsection{Identifiability Considerations}

The MSFR model, like other standard factor analysis models, is non-identifiable up to orthogonal transformations  and sign switches of the loadings (common $\bsp$ and study specific $\bsL_s$). Although this is a mathematical constraint, it does not cause any issues in real data analysis. Specifically, for interpretability, the individual signs of the entries in the loading column are not important; only the contrast of their signs matters for naming the factors (Sections~\S~4 and 5).
In addition to those sources of non-identification, the MSFR model also exhibits indeterminacy in the decomposition of the covariance matrix $\bsS_s$.

The rotation identifiability arises since any orthogonal transformation of the factors and the loading matrices, common and study-specific, leads to the same distribution of $\mbx_{is}$. 
Thus, the marginal distribution of $\mbx_{is}$ can equivalently be redefined as a multivariate normal with covariance matrix $\bsS_s =\widetilde{\bsp} \widetilde{\bsp}^\top + \widetilde{\bsL}_s \widetilde{\bsL}_s^\top + \bsP_s$, where $\widetilde{\bsp} = \bsp \mbR$ and $\widetilde{\bsL}_s = \bsL_s \mbR_s$, with $\mbR \in \R^{q \times q}$ and $\mbR_s \in \R^{q_s \times q_s}$ orthogonal matrices.
Furthermore, Model~\eqref{eq:model} can be rewritten as $\mbx_{is} =  {\bsb} \mbb_{is} + \widetilde{\bsp} \widetilde{\mbf}_{is} +   \widetilde{\bsL}_s \widetilde{\mbl}_{is}  + \mbe_{is}$, with $\widetilde{\mbf}_{is} = \mbR^\top {\mbf}_{is}$ and $\widetilde{\mbl}_{is} = \mbR_s^\top \mbl_{is}$.

In the factor analysis literature, several strategies have been developed to tackle this issue, such as performing a varimax rotation \citep{Kaiser1958Tvcf}, or restricting the loading matrix to be lower triangular \citep{Lopes2004BMAI, Carvalho2008HSFM} or generalized lower triangular \citep{FruehwirthSchnatter2023SBfa}. 
Here, we extend the traditional varimax rotation to both the $\bsp$ and $\bsL_s$, which was selected for its interpretability.

Under varimax rotation and sign constraints, a full rank loading matrix is uniquely determined in regular factor analysis, giving the error term.
However, in multi-study latent models, an additional source of non-identifiability arises, the indeterminacy of the covariance $\bsS_s$ decomposition.
Specifically, the difference between the covariance matrix and the idiosyncratic precisions can be redefined as $\bsS_s - \bsP_s ={\bsp^*} {\bsp}^{*\top} + {\bsL}^*_s {\bsL}_s^{*\top}$, where the column vectors of $\bsp$ and ${\bsL}_s$ can move from one loading matrix to the other. 
To solve this issue, we follow the strategy of \citep{DeVito2019Mfa}, and constrain the matrix $[\bsp, \bsL_1,\dots, \bsL_S] \in \R^{p \times (q + \sum_{s=1}^S q_s)}$ matrix, containing both the common and study specific loadings, to have a full column rank $q + \sum_{s=1}^S q_s \leq p$. 
Thus, the span of the column vectors of
$\bsS_s - \bsP_s$ is equal to the sum of the span of the column vectors of $\bsp$ and ${\bsL}_s$, leading to uniquely determined loading matrices.

\section{Simulation Study}
\label{sub:Simulations}
We first use simulation studies to assess the ability of our MSFR model to detect the true signals for both the common and the study-specific and outperform the standard MSFA in the presence of observed covariates. Specifically, we compare the MSFR with MSFA  to illustrate how much inference can suffer when not accounting for covariate effects.
To evaluate if the covariate effect can be removed from the data and then adopt the MSFA to the residuals, we fit a linear regression (LR) model, followed by MSFA estimation on the residuals, here called (MSFA\&LR). 
Moreover, we compare the MSFR with the vanilla Factor Regression (FR) \citep{Avalos2022HLDI, Carvalho2008HSFM} to assess the advantages of estimating the study-specific factors. 
The R-code for our methods is available at \texttt{\textbf{https://github.com/rdevito/MSFR}}. 

We consider various scenarios accounting for different sample sizes $n_s$, number of studies $S$, number of observed variables $p$, structure of $\bsb$, and structure of $\bsS$. 
Scenario 1 shows the advantages of our method in situations with small covariate effects. 
It assumes $S=2$ studies, each with a sample size of $n_s=500$ and $p=20$ responses.
This scenario assumes $q=3$ common factors, $q_s =1$ study-specific factor, and $p_b = 2$ covariates.
Scenarios 2 and 3 closely mimic the case study data, the HCHS / SOL data set, that is, the same number of studies and patient characteristics. 
These scenarios are designed to evaluate the performance of our method in more complex settings with larger datasets and multiple covariate effects.
Scenarios 2 and 3 assume $q=4$ common factors, $q_s=1$ study-specific factors, and $S=6$ studies.
Each study has $p = 42$ responses, with varying sample sizes $n_s = \{1257,1444,2126,4940,2314,897 \}$.  
The main difference between Scenario 2 and 3 lie on the structure of $\bsb$. Specifically, Scenario 2 assumes that $\bsb$, $p_b = 7$, is common to the six studies, while Scenario 3 assumes that each study has a study-specific $\bsb$, $p_b = 6$, here column $\bsb_{\cdot s}, s=1,\dots,6$ represents the study-specific covariate effect.

For all the scenarios, we simulated 100 collections of datasets from $N(\bsb \mbb_{is},\bsS_s)$, with $\bsS_s=(\bsp \bsp^\top +\bsL_s \bsL_s^\top + \bsP_s)$. 
The common $\bsp$ and study specific $\bsL_s$ loading matrices are assumed to be sparse, with one-third of non-zero entries, randomly generated from the uniform densities $\bsp \sim \text{sign}\times U(0.6,1)$, with $\text{sign}\sim U\{-1,1\}$ and $\bsL_s \sim U(-1,1)$. 
Finally, the idiosyncratic precision matrices, $\bsP_s$, were assumed to be diagonal matrices with the entries generated uniformly between (0,1).

\begin{figure}[!ht]
\centering
\includegraphics[scale=.315]{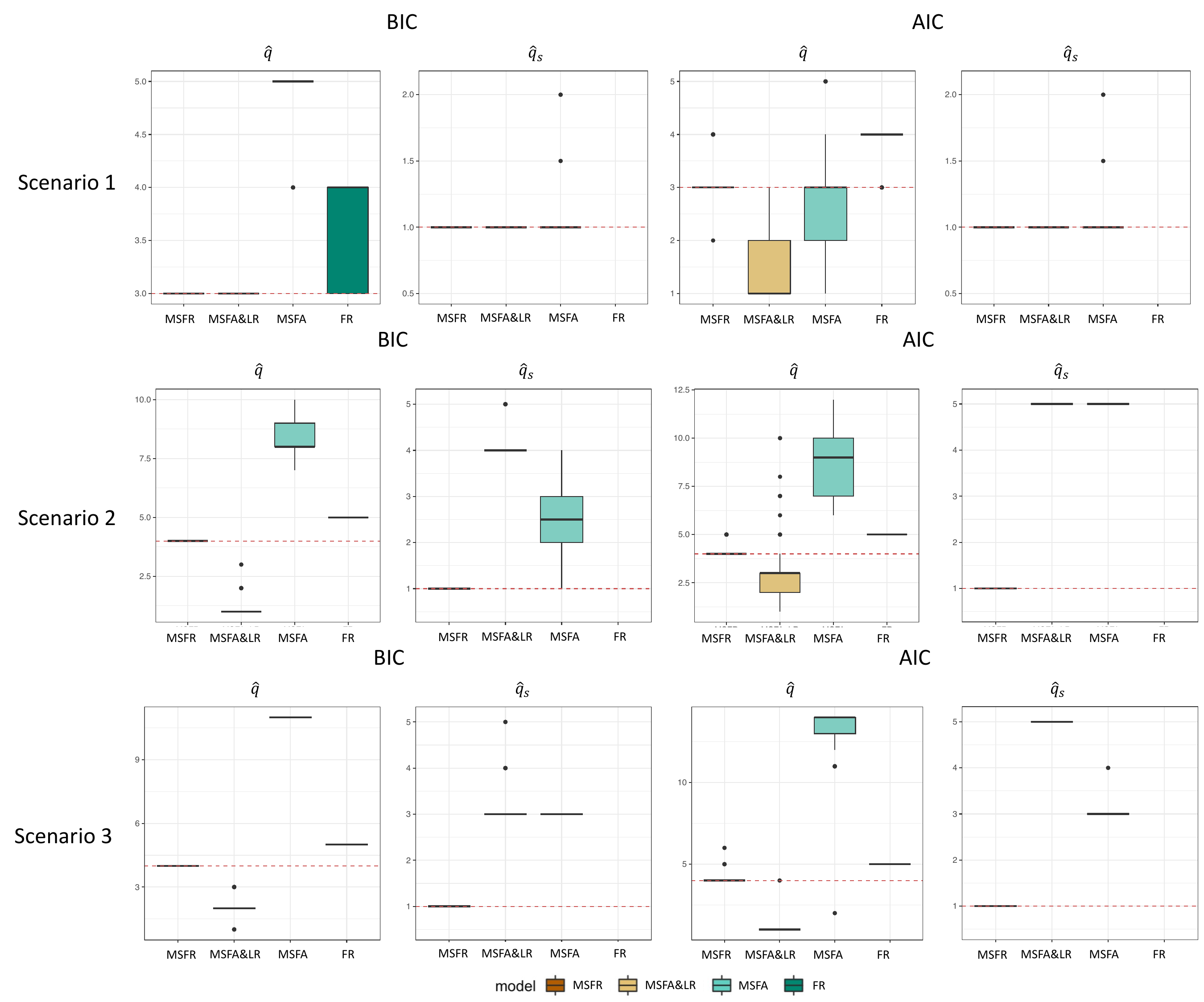}
\caption{Boxplots of the common and study-specific factor cardinality reconstructions, $\widehat{q}, \widehat{q}_s$ respectively, across 100 simulations, based on the BIC and AIC criteria for the three different scenarios and all models. The red dotted lines represent the ground truth cardinalities.}
\label{fig:BoxCardinality}
\end{figure}

We first evaluate the MSFR's ability
to recover the common and study-specific latent dimension, respectively $\widehat{q}$  and $\widehat{q}_s$  (Figure~\ref{fig:BoxCardinality}). We show the mean results across 100 simulations of  $\widehat{q}$ and $\widehat{q}_s$. Our method is consistently  able to estimate the right number of dimension in all scenarios' settings,  independently of the model selection criteria (BIC and AIC), showing a most robust behaviour than MSFA\&LR and MSFA. As expected, FR required a higher number of factors to take into account the impact of the study-specific factors. Likewise, MSFA required more factors since it does not consider the covariate effects. The MSFA\&LR showed the worst performance in the reconstruction of $\widehat{q}$, $\widehat{q}_s$, 
highlighting the limitations of using two-step procedures relative to a joint estimation of the multiple-factor model and covariate effects. 

\begin{figure}[!ht]
\centering
\includegraphics[scale=.39]{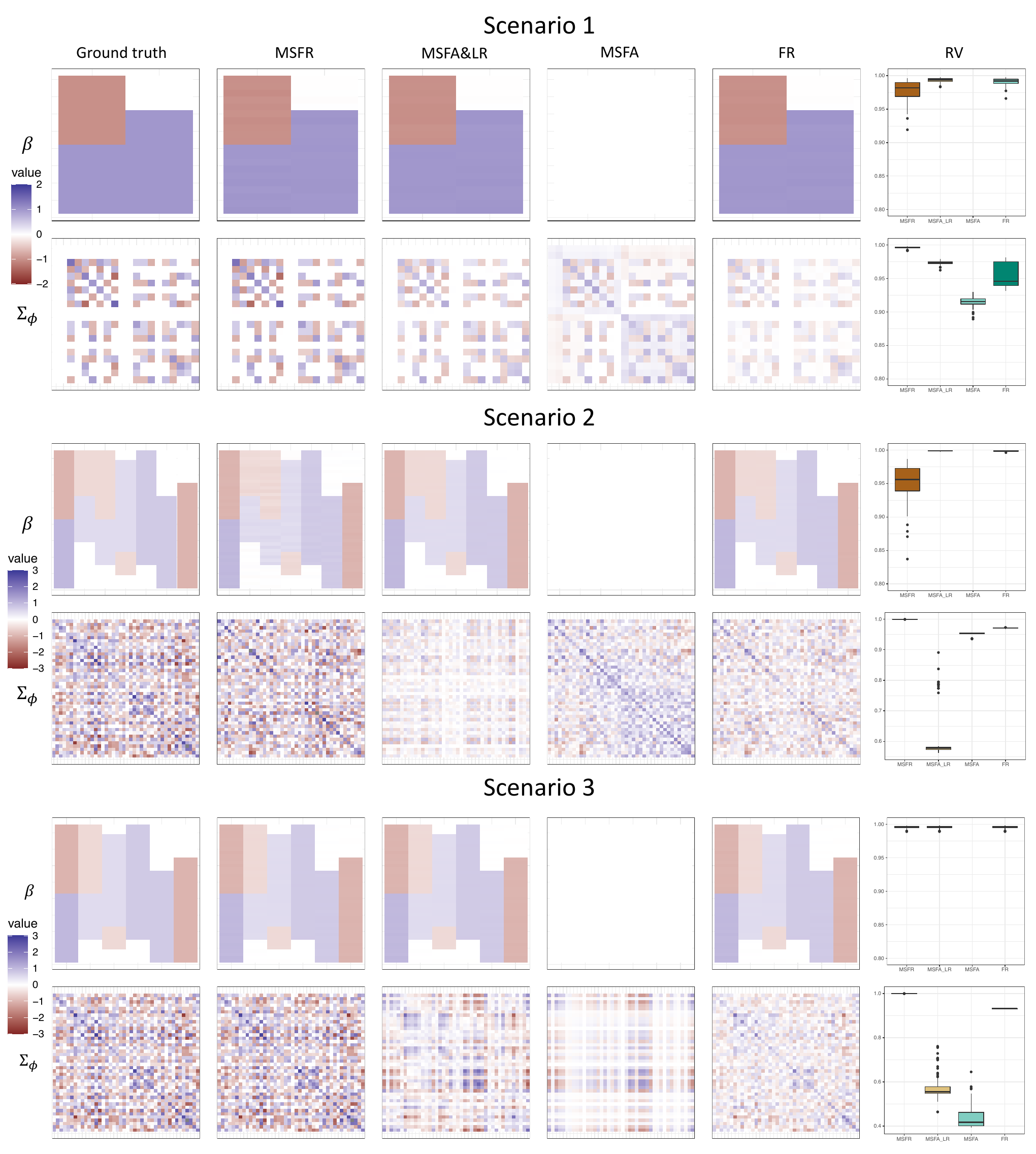} 
\caption{Heatmaps of the average values of the true $\bsb$ and $\bsS_{\bsp}$, along with their average estimates for all four methods (left panels), and the boxplots of their RV coefficients (right panel). Results are based on 100 simulations, with models selected using the BIC criterion for all three scenarios.}
\label{fig:HeatAll}
\end{figure}

\begin{figure}[!ht]
\centering
\includegraphics[scale=.425]{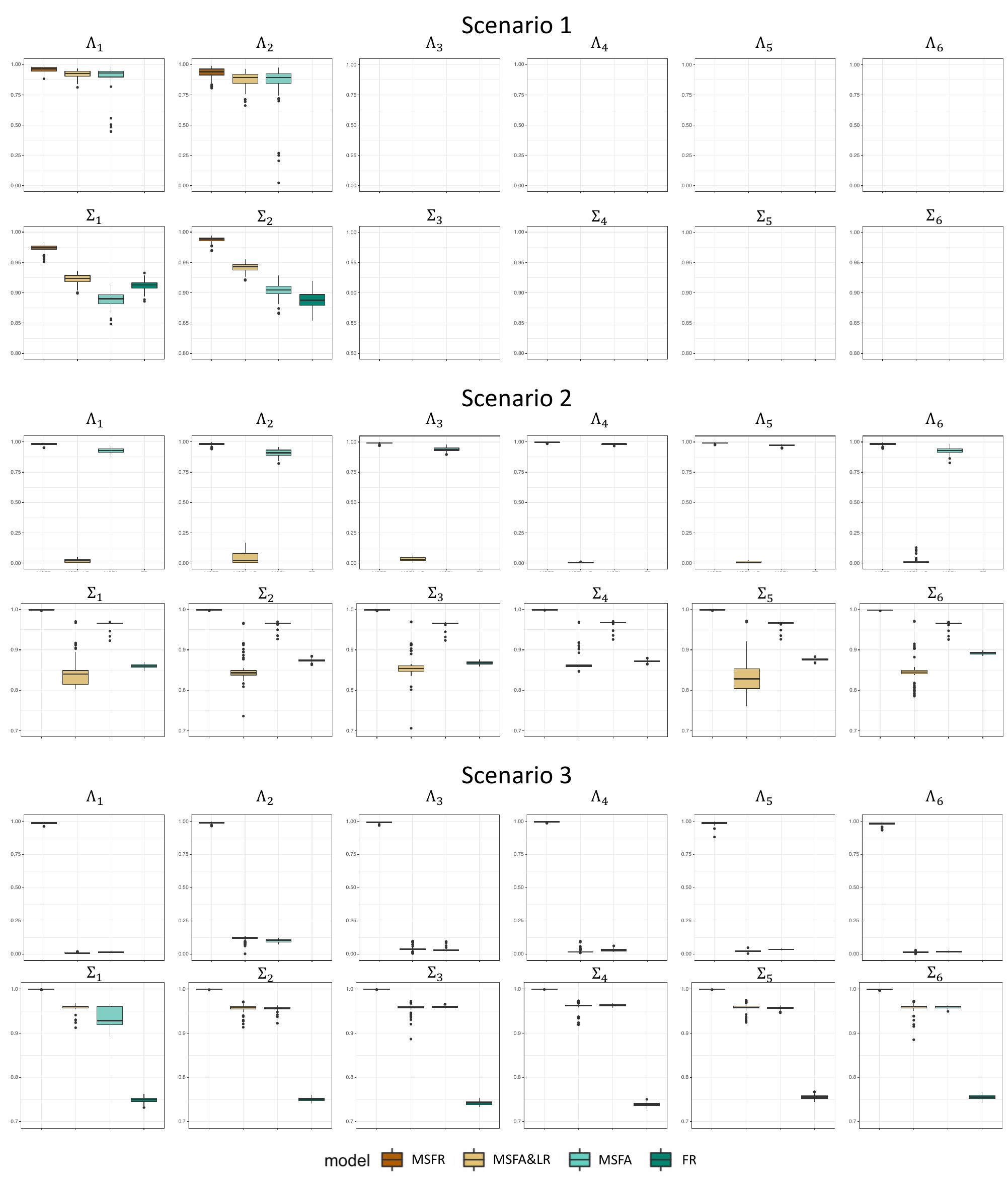} 
\caption{Boxplots of the RV values of the estimated $\bsl_{s}$ and $\bsS_s$ for all the three simulated scenarios and all the models. Results are based on 100 simulations, with models selected using the BIC criterion.}
\label{fig:BoxPlotsAll}
\end{figure}

We then proceed on the recovery of  the covariance component $\bsS_\Phi$ determined by the common factors, as well as the beta coefficient for the covariates $\widehat{\bsb}$.
To measure the similarity between $\bsS_\Phi$ and $\widehat{\bsS}_\Phi$ we use the RV coefficient \citep{robert1976}.
The RV coefficient is a measure of similarity between two matrices $\mathbf{A}$ and $\widehat{\mathbf{A}}$ varying between 0 and 1 defined as 
\begin{equation}\label{eq:RV-def}
    \text{RV}(\mathbf{A},\widehat{\mathbf{A}}) = \frac{\text{Tr}(\mathbf{A}\widehat{\mathbf{A}}^\intercal\widehat{\mathbf{A}}\mathbf{A}^\intercal)}{\sqrt{\text{Tr}(\mathbf{A}\mathbf{A}^\intercal)^2\text{Tr}(\widehat{\mathbf{A}}\widehat{\mathbf{A}}^\intercal)^2}}.
\end{equation}
The RV coefficient varies in $[0, 1]$, the closer RV is to 1, the more identical the two matrices are \citep{robert1976}.
We would like to highlight that the RV coefficient provides the same value to distinct loading matrices that differ by an orthogonal rotation. 
This is because it takes into account $\widehat{\mathbf{A}}\widehat{\mathbf{A}}^\intercal$, which is invariant to orthogonal transformations.
For each simulation, we considered a grid of possible values for $\widehat{q}$ and $\widehat{q}_s$, and
we fit the models for all the grid values and performed model selection. 
Figure~\ref{fig:HeatAll} 
displays the heatmaps of the true $\bsb$ and $\bsS_{\bsp}$, and their average estimations (left panel), as well as the RV's box-plots for $\bsb$ and ${\bsp}$ (right panel) using the BIC approach for all the 3 scenarios. 
The results showed the advantages of using our proposed method. 
MSFR presented the highest RV for the common and study-specific loading matrices as well as for the whole covariances. 
Figure~
\ref{fig:BoxPlotsAll} 
shows the box-plots of the RV coefficients of $\bsl_{s}$ and $\bsS_s$ for all the three scenarios.
We remark that the panels for \(\bsl_3, \ldots, \bsl_6\) and \(\bsl_3, \ldots, \bsl_6\) are blank in Scenario 1 because it only involves two studies, and hence, these variables are not infer in this case.
Overall,  MSFR performs better than the other methods, with the smallest variance in all  cases, except for $\bsb$ 
Figure~\ref{fig:HeatAll}). 
It is important to highlight that even though MSFA achieved a competitive reconstruction of $\bsp$, the  average $\bsS_{\bsp}$ estimates over the simulations are less precise, as displayed in the heatmaps in Figure~\ref{fig:BoxPlotsAll}. 

Table \ref{tab:RunningTimes} presents the time required to run the simulations across all scenarios. Since our approach is model-based, it simultaneously corrects for covariate adjustments while learning both the common and study-specific factors and loadings. As a result, it is the most computationally intensive method compared to our competitors. Nevertheless, in our most complex scenarios (2 and 3), which closely mimic the case study data, the procedure only took approximately 6 minutes to generate inferences. 

Finally, Supplementary C, Figures~S1 and S2 present the results of the reconstructions of $\bsb$, $\bsS_{\bsp}$, $\bsl_{s}$ and $\bsS_s$ obtained via the AIC criterion. 
The figures demonstrate consistent patterns and results for the MSFR compared to the ones obtained via the BIC procedure. 
However, the results showed that the performance of our competitors varied significantly, further corroborating that those methods may not be robust to the model selection criteria.

\begin{table*}[t]
\centering
\caption{\it  Running times in seconds over 100 simulations}
\label{tab:RunningTimes}
 \tabcolsep=0pt
    \begin{tabular*}{\textwidth}{@{\extracolsep{\fill}}c|cc|cc|cc@{\extracolsep{\fill}}}
&\multicolumn{2}{c|}{\textbf{Scenario 1}: $q=3, q_s=1$} &  \multicolumn{2}{c|}{\textbf{Scenario 2}: $q=4, q_s=1$} & \multicolumn{2}{c}{\textbf{Scenario 3}: $q=4, q_s=1$} \\
\hline
& mean & sd  & mean & sd & mean & sd \\ 
\hline
  \textbf{MSFR} & 32.63 & 2.38 & 376.74 & 36.50 & 364.45 & 30.99 \\ 
  \textbf{MSFA\&LR} & 12.61 & 1.09 & 39.98 & 2.19 & 40.74 & 3.28 \\ 
  \textbf{MSFA} & 14.98 & 1.15 & 113.15 & 7.12 & 193.94 & 14.32 \\ 
  \textbf{FR} & 0.96 & 2.16 & 6.60 & 1.28 & 6.45 & 1.13 \\ 
 \hline
 \hline
\end{tabular*}
\end{table*}


\section{The HCHS/SOL application}
\label{sub:HCHS/SOL}

We demonstrate the usefulness of our method by performing three different analyses on the HCHS/SOL.
First, we compare the prediction accuracy between the MSFR and the MSFA to show the power of our method in predictive ability.
Second, we explore the common and study-specific factors loadings among the five different ethnic background groups of the HCHS/SOL.
Third,  we even validate our estimated dietary patterns with diabetes, high cholesterol and hypertension, three of the strongest risk factors for almost all different cardiovascular diseases, including cerebral stroke and renal failure \citep{kjeldsen2018hypertension}.

Moreover, before to proceed, we calculated the BIC for various models, each incorporating different combinations of covariates and factors, to determine the best-fitting model (Table S4, Supplementary Materials). The final model, presented below, was selected based on the minimum BIC.

\textbf{Predictive analyses.} 
We compare the k-fold cross-validation (setting k=5) prediction errors computed by our MSFR and the MSFA. Thus, we fit both the MSFR and the MSFA model on a random 80\% of the data, followed by a different 80\% for different folds. We then evaluate the prediction error on the remaining 20\% in each fold. Predictions are performed as
$$
\mbox{MSFR: } \widehat{\mbx}_{is} = \widehat{\bsb} \widehat{\mbb}_{is} + \widehat{\bsp}^{MSFR} \widehat{\mbf}_{is} + \widehat{\bsL}^{MSFR}_s \widehat{\mbl}_{is}  $$
$$
\mbox{ MSFA: } \widehat{\mbx}_{is} =  \widehat{\bsp}^{MSFA} \widehat{\mbf}_{is} + \widehat{\bsL}^{MSFA}_s \widehat{\mbl}_{is},
$$
where $\mathbf{x}_{is}$ is the vector of foods consumption for individual $i$ in study $s$, $\widehat{\bsp}^{MSFR}$ is the common factor loadings obtained with the MSFR method, while    $\widehat{\bsp}^{MSFA}$ is the common factor loadings obtained with the MSFA method; $\widehat{\bsL}^{MSFR}_s$ is specific factor loadings estimated with MSFR, while $\widehat{\bsL}^{MSFA}_s$ is specific factor loadings.
The factor scores are then computed using the Bartlett method since scores are
uncorrelated with each other \citep{distefano2009understanding},  
see Supplement~D for further details, by using different parameters obtained in the MSFR and MSFA. 
We then assessed the mean squared error (MSE) of prediction for each fold as
$
MSE =\frac{1}{n} \sum_{s=1}^S \sum_{i=1}^{n_s} (\mbx_{is} - \widehat{\mbx}_{is})^2,
$
using the 20\% of the samples in each study set aside for each cross-validation iteration. We averaged all the MSE obtained in each fold. Table~\ref{tab:MSE} shows that the obtained MSE obtained with our  MSFR method is 1.45  and the MSE obtained with the MSFA method is 2.34, thus the MSE is 61\% smaller for MSFR than for MSFA. 
This analysis reveals how MSFR borrows strength across studies in estimating the factor loadings and clearly detecting factor loading signals by removing all the covariates noise that can impact them. Thus, our predictive power in independent observations is not only preserved but also greatly improved.  

\begin{table*}[]
\centering
\caption{Mean square error (MSE) obtained via k-fold cross validation for both the multi-study
factor regression model (MSFR) and the multi-study factor analysis model (MSFA).}
\begin{tabular}{ccc}
\textit{Method} & {\textit{MSE}} &\textit{Relative Efficiency}\\
\hline
\hline
\textbf{MSFR}  & 1.45 & 1.00 \\
\textbf{MSFA}  & 2.34 & 1.61 \\
\hline
\end{tabular}
\label{tab:MSE}
\end{table*}

\textbf{Factor loadings analyses.}
We considered two main tasks: to analyze the estimated common and study-specific factors loadings, $\bsp$ and $\bsL_s$, respectively.

The MSFR model selected $\widehat{q} = 5$ common factors and $\widehat{q}_s = 1$ study-specific factor for the six different ethnic backgrounds (obtained with the AIC procedure).

\begin{figure*}[!ht]
\caption{\it Heatmap of the common (left panel) and ethnic background-specific (right panel) factor loadings in the HCHS/SOL. Loadings in blue (orange) represent positive (negative) associations between factors and foods.}\label{fig:com}
\includegraphics[width=\textwidth]{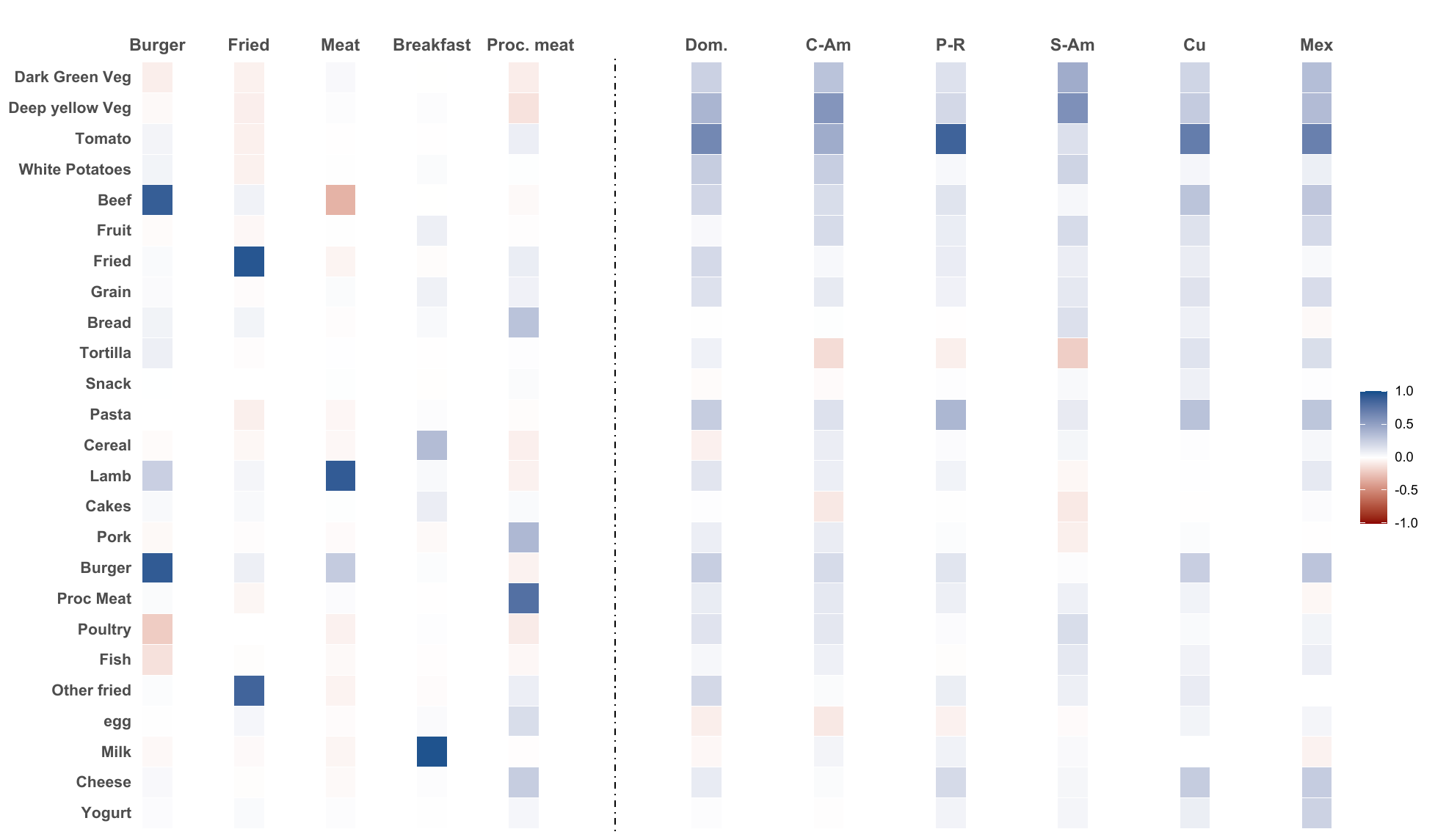} 
\centering

\end{figure*}

We proceed by inspecting the factor structures.  Figure~\ref{fig:com} provides a visual representation of the common (right panel) and study-specific loadings (left panel). 
In order to facilitate interpretation and thus name each factor, we performed a varimax rotation \citep{Kaiser1958Tvcf} for the common loading matrix. This rotation was unnecessary for $\bsL_s$ as our MSFR estimates only one extra study-specific loading. 

The varimax rotation in the common loadings maximizes the variance in the squared rotated loadings, making the loadings extreme and positive for a better interpretation.
Factor 1, namely Burger, 
showed high-valued loadings in Beef and Burger, while negative loadings in Poultry and Fish. 
Factor 2 displayed high-valued loadings in Fried (potato chips, plantains, etc.) and Other Fried ( Chicken nugget, etc.) and thus was named Fried. Factor 3, namely Meat, 
exhibited high-valued loadings on Lamb and Burger but is negative in Beef, for this reason was named Lamb Meat. Factor 4, namely Breakfast, 
presented high-valued loadings on Milk, Cereal, and medium loadings on Grain and Fruit. Finally factor 5, namely Processed Meat had high loadings on Processed Meat, Pork, and Bread. Moreover, to quantify uncertainty, we perform a bootstrapped analysis. Specifically, we compare the distributions of the shared factor loadings computed from 100 bootstrapped random sets of our original six ethnic background groups under our model, the MSFR, and the MSFA (Figure~S3, Supplementary Materials). The boxplots representing the distribution of the loadings across the 100 bootstrapped random sets were, in the majority of the cases, narrower under our MSFR model than under the MSFA method, which even exhibited strong outliers. This suggests that incorporating confounders/covariates in the model can make the estimates more stable.

In the absence of any rotation in the specific factor loadings, we based our interpretation on the opposite (positive and negative) signs or colors (blue and dark red, respectively) illustrated in the study-specific factor loading heatmap (Figure~\ref{fig:com} right panel). 
They displayed variations, showing heterogeneity across ethnic background cultures and providing a deeper understanding of the ethnic background specific dietary patterns. 
There is a general trend since all the groups eat tomatoes. However, while the Dominican group seems to eat tomatoes with vegetables (Deep Yellow and Dark Green Veg), the Puerto Rican and Cuban groups eat tomatoes with pasta and cheese. The South American group is an extremely vegetarian group, eating more Dark Green and Deep Yellow Veg than other food groups.
This is further corroborated by the factor congruence, a measure of similarities of factor loadings, closer to 1 (0) higher is similarities (dissimilarities). The greatest factor congruence is between Puerto Rican and Cuban (0.8), while the others are around 0.23 showing very poor similarities.

\textbf{Association with diabetes high cholesterol and hypertension.} 
In order to validate and associate each factor with diabetes, high cholesterol, and hypertension from 24-hour recalls, we estimate the factor score by adopting the Thurstone method in the multi-study setting. Factor scores reveal the degree to which each subject's diet adheres to one of the identified factors.

We first divided each common factor score into quartile categories and each ethnic background specific score into tertile categories.
We then used these quantile-based categories to validate each estimated factor with diabetes, high cholesterol, and hypertension provided by HCHS/SOL. 

We performed a generalized weighted logistic regression model \citep{lumley2017fitting}, incorporating the sample weight provided by HCHS/SOL. We remark that we have not included any covariates as confounders since we have already accounted for those in our multi-study factor regression model.

\begin{figure*}[ht!]
\caption{\it  Odds ratio and 95\% confidence intervals of Diabetes, High cholesterol and Hypertension on the five common factors for the quartile categories, with the first quartile as the baseline. Estimation are obtained from a weighted logistic regression.}
\includegraphics[width=\textwidth]{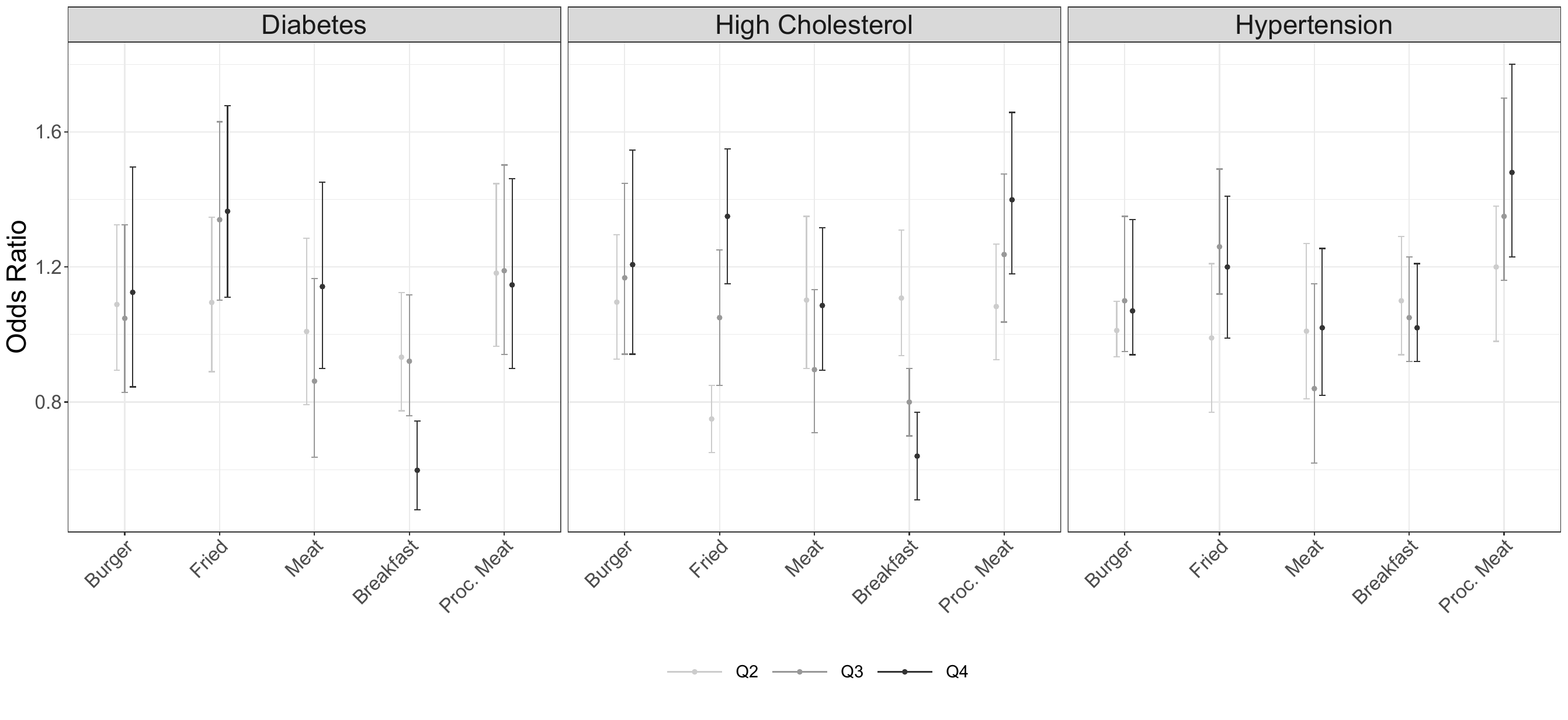} 
\label{fig:fg_comRisk}
\centering
\end{figure*}

For the common dietary patterns (Figure~\ref{fig:fg_comRisk}), all the patterns increase the risk of diabetes, high cholesterol and hypertension, except for the Breakfast pattern. Specifically, the breakfast pattern significantly reduces the risk of diabetes (OR=0.6) and high cholesterol (OR=0.68) while not having any effects on hypertension (OR=1.01). In the breakfast pattern, there are foods like cereal, oatmeal, wholegrain wheat breakfast cereal biscuits, and high-fiber cereal that can significantly reduce the LDL-cholesterol \citep{davy2002high, clifton2018cholesterol}. The fried pattern significantly increases both the risk of diabetes and high cholesterol, while the Processed meat is significantly associated with the probability of high cholesterol and hypertension.

\begin{figure*}[ht!]
\caption{\it  Odds ratio and 95\%  confidence intervals of Diabetes, High cholesterol, and Hypertension on each ethnic background-specific factor for the tertile categories, with the first tertile as the baseline.  Estimation are obtained from a weighted logistic regression.}
\includegraphics[width=\textwidth]{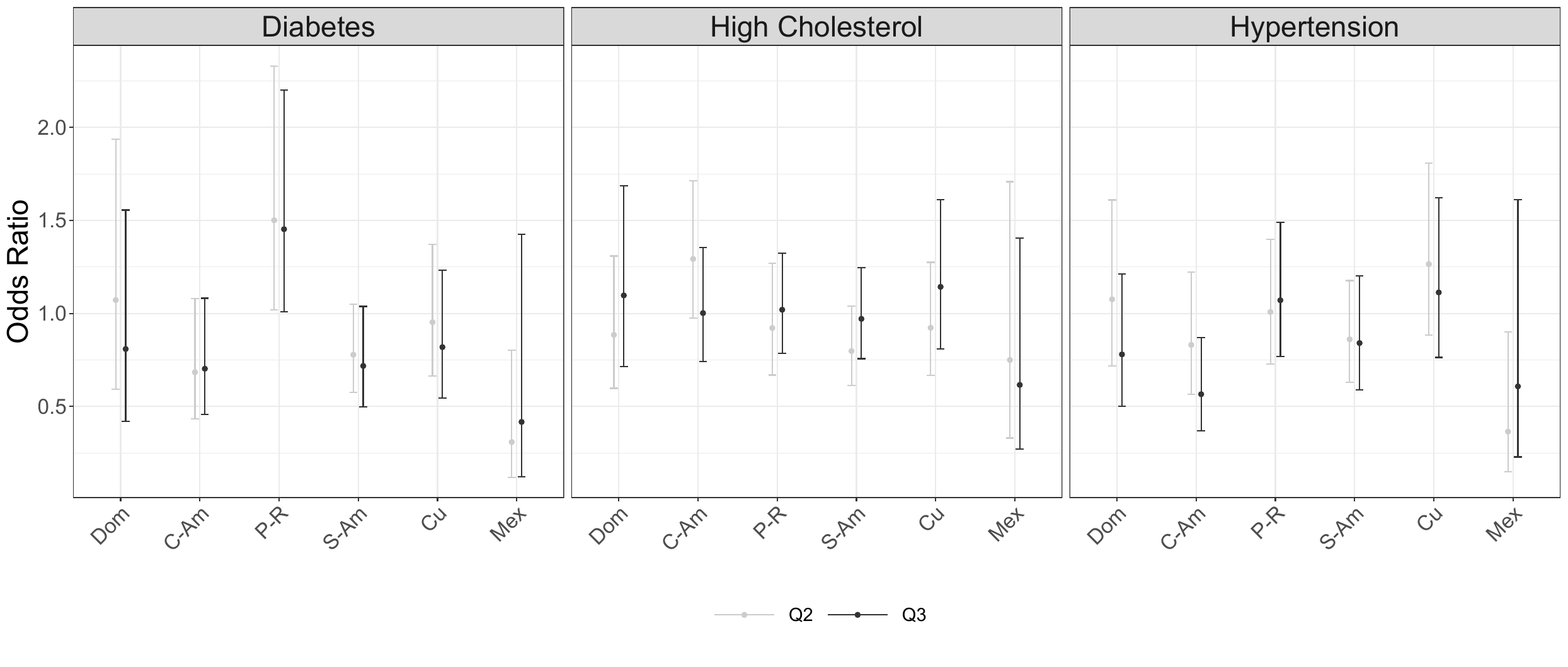} 
\label{fig:fg_specRisk}
\centering
\end{figure*}

The ethnic background-specific dietary patterns reveal different associations with health outcomes  (Figure~\ref{fig:fg_specRisk}. Only the Puerto-Rican dietary pattern significantly increases the risk of diabetes (OR: 1.495, 95\%CI: 1.01- 2.20); all the other patterns either show no significant influence on diabetes or appear protective (i.e., the Mexican dietary pattern, OR: 0.46, 95\% CI: 0.27-1.49).  For high cholesterol, no significant associations were observed across the dietary patterns. However, we observe that only the Cuban and Dominican dietary patterns increase the risk of high cholesterol. Conversely, only the Puerto-Rican and Cuban dietary patterns increase the probability of hypertension in the highest tertile of consumption. 
These two ethnic background dietary patterns are characterized by higher consumptions of pasta and cheese compared to the other patterns (i.e., associated with consumption of vegetables and tomatoes). In fact, high consumption of pasta and cheese has been found to be associated with diabetes, hypertension, and cardiovascular risk factors \citep{vitale2019pasta, glazener1964pargyline}.

These analyses highlight that the MSFR is able to capture dietary patterns shared across all studies and also reveal new insight into association disease analysis.

We further demonstrate the usefulness of our method with the HCHS/SOL nutrients dataset in Section E of the supplementary material. 

\section{Discussion}

In this article, we aim to identify the shared and ethnic background dietary patterns in the HCHS/SOL study, while adjusting for socio-demographic characteristics or confounders that do not affect the dietary patterns, such as gender, education level, alcohol and cigarette intake, and BMI. 
To achieve this goal, we have developed a novel class of factor models: the Multy-Study Factor Regression (MSFR). 
In this article, we aim to identify the shared and ethnic background dietary patterns in the HCHS/SOL study, while adjusting for socio-demographic characteristics or confounders that do not affect the dietary patterns, such as gender, education level, alcohol and cigarette intake, and BMI. To achieve this goal, we have developed a novel class of factor models: the Multi-Study Factor Regression (MSFR).

Our MSFR integrates data from multiple sources using joint dimension reduction, via Multi-study Factor Analyses \citep{DeVito2019Mfa, DeVito2021Bmfa}, and covariate effect adjustment, via latent factor regression \citep{Avalos2022HLDI, Carvalho2008HSFM}. 
Our model is fitted deterministically using an efficient ECM algorithm with closed-form updates, and it is available publicly at \texttt{\textbf{https://github.com/rdevito/MSFR}}.

Bayesian approaches for standard factor analysis regression and the MSFA have been explored \citep{Carvalho2008HSFM, Avalos2022HLDI, DeVito2021Bmfa}. Thus,  adopting a Bayesian framework for MSFR could be a promising direction for high-dimensional data, particularly in $p > n$ settings. Bayesian techniques are particularly beneficial in such settings, like gene expression or somatic mutation data, though they often rely on Markov Chain Monte Carlo (MCMC), which is computationally demanding. In this study, we adopt an EM algorithm, achieving point estimates similar to variational inference \citep{hansen2024fast} but with much lower computational overhead and greater interpretability. In fact, Bayesian approaches require additional post-processing steps to resolve identifiability issues across each chain. In such scenarios, identifiability constraints may need to be revisited, as it may not be necessary to enforce \([\Phi, \Lambda_1, \ldots, \Lambda_S]\) to be of full rank. Additionally, the lower triangular constraint should be avoided, as an adequate number of zeros in the structure can prevent orthogonal transformations \citep{Schnatter}. Further research is needed to explore whether these adjustments would remain effective in the context of multi-study factor analysis.

Moreover, methods such as the EM algorithm and variational inference provide point estimates of the unknown parameters, without offering any quantification of uncertainty. Methods such as bootstrapping can alleviate this limitation. In our analysis, we have provided bootstrapped histograms for each shared factor in the real data analysis, as shown in Figure S3 (Supplementary Materials).

Our proposed MSFR has the potential to be adopted in several different data settings where the goal is to disentangle the similarities and dissimilarities among heterogeneous populations, studies or groups, while adjusting for covariate effects. The covariate adjustment enables MSFR to better capture common and study-specific loadings and latent factors. Moreover, our model handles covariates that can either be the same or different across studies. The MSFR addresses this complexity by allowing it to account for variations in both the response variables and covariates across different populations and generalize more effectively. The proposed covariate adjustment essentially uses information about relationships between socio-demographic characteristics and the outcome so that MSFR can better identify the common and study-specific loading matrices and latent factors. 
Additionally, the populations differ, meaning the units vary across studies. A potential future direction could involve scenarios where some units are shared between two or more studies, which would create challenges in maintaining uniformity in the response variables and thus assuming a dependent structure in the likelihood.

Our model outperforms its competitors, MSFA and FR, in estimating factor cardinality (both common and study-specific), with MSFA requiring more factors to handle covariate effects and the absence of study-specific factors.

In our nutritional epidemiology application, we show that the covariate adjustment of our MSFR improves the quality of our estimations and predictions. 
Furthermore, our model was able to provide interpretable dietary patterns for the whole Hispanic
Community, as well as for each Latino ethnic background. 
The factors obtained, both common and ethnic background-specific, were explained and named according to their food group values. In latent factor models, the interpretation of factors remains a key challenge, as there is no standard method for naming them. For example, in big-data genomic scenarios, gene set enrichment analysis (e.g., \citep{tyekucheva2011integrating}) is a post-hoc approach that determine whether each latent factor is significantly associated with a well-known biological pathway \citep{de2022shared}. In nutritional epidemiology, however, it is standard practice to name factors based on their highest loadings, as we have done in this study.
Moreover, we illustrated the utility of our method in a supervised framework by studying the factor association with diabetes, high cholesterol and hypertension. 
We emphasize that our proposed method eliminates the need for an additional covariance adjustment when analyzing  the relationships with high cholesterol and hypertensio, which is required in the case of MSFA.

We remark that our novel MSFR can also be extended beyond nutritional epidemiology, for example, to cancer genomics, to identify patterns of common and study-specific transcriptional variation while adjusting for pre-treatment patient characteristics, or in finance to study common market trends and specific economic cycles, while including other economic variables.

In summary, MSFR is essential in the case application considered here. We hope it will help the joint analyses of multiple studies in many different settings,  abating the current challenges of reproducibility of unsupervised analyses in nutritional epidemiology and data science.

\section{Competing interests}
No competing interest is declared.

\section{Acknowledgments}
This Manuscript was prepared using HCHSSOL Research Materials obtained from the NHLBI Biologic Specimen and Data Repository Information Coordinating Center and does not necessarily reflect the opinions or views of the HCHSSOL or the NHLBI. 
The authors would like to thank David Rossell and Giovanni Parmigiani for their precious advice, as well as Blake Hansen for his assistance in deploying the code to the server.

\bibliography{regMSFA}

\begin{thebibliography}{60}
\providecommand{\natexlab}[1]{#1}
\providecommand{\url}[1]{\texttt{#1}}
\expandafter\ifx\csname urlstyle\endcsname\relax
  \providecommand{\doi}[1]{doi: #1}\else
  \providecommand{\doi}{doi: \begingroup \urlstyle{rm}\Url}\fi

\bibitem[Willett and MacMahon(1984)]{willett1984diet}
Walter~C Willett and Brian MacMahon.
\newblock Diet and cancer-an overview.
\newblock \emph{New England Journal of Medicine}, 310\penalty0 (11):\penalty0 697--703, 1984.

\bibitem[Kromhout(2001)]{kromhout2001diet}
Daan Kromhout.
\newblock Diet and cardiovascular diseases.
\newblock \emph{The journal of nutrition, health \& aging}, 5\penalty0 (3):\penalty0 144--149, 2001.

\bibitem[{GBD 2013 Risk Factors Collaborators and others}(2015)]{gbd2015global}
{GBD 2013 Risk Factors Collaborators and others}.
\newblock Global, regional, and national comparative risk assessment of 79 behavioural, environmental and occupational, and metabolic risks or clusters of risks in 188 countries, 1990--2013: a systematic analysis for the global burden of disease study 2013.
\newblock \emph{Lancet}, 386\penalty0 (10010):\penalty0 2287, 2015.

\bibitem[Aune et~al.(2016)]{aune2016nut}
Dagfinn Aune et~al.
\newblock Nut consumption and risk of cardiovascular disease, total cancer, all-cause and cause-specific mortality: a systematic review and dose-response meta-analysis of prospective studies.
\newblock \emph{BMC medicine}, 14\penalty0 (1):\penalty0 1--14, 2016.

\bibitem[nat(Accessed May 3, 2024)]{nat_2024}
Multiple cause of death 2018–2022 on {CDC WONDER} database.
\newblock \emph{National Center for Health Statistics}, Accessed May 3, 2024.

\bibitem[Stearns et~al.(2017)]{stearns2017ethnic}
Jennifer~C Stearns et~al.
\newblock Ethnic and diet-related differences in the healthy infant microbiome.
\newblock \emph{Genome medicine}, 9:\penalty0 1--12, 2017.

\bibitem[Zulyniak et~al.(2017)]{zulyniak2017does}
Michael~A Zulyniak et~al.
\newblock Does the impact of a plant-based diet during pregnancy on birth weight differ by ethnicity? a dietary pattern analysis from a prospective canadian birth cohort alliance.
\newblock \emph{BMJ open}, 7\penalty0 (11):\penalty0 e017753, 2017.

\bibitem[Haines et~al.(1999)Haines, Siega-Riz, and Popkin]{haines1999diet}
Pamela~S Haines, Anna~Maria Siega-Riz, and Barry~M Popkin.
\newblock The diet quality index revised: a measurement instrument for populations.
\newblock \emph{Journal of the American Dietetic Association}, 99\penalty0 (6):\penalty0 697--704, 1999.

\bibitem[Schulze et~al.(2003)Schulze, Hoffmann, Kroke, and Boeing]{schulze2003approach}
Matthias~B Schulze, Kurt Hoffmann, Anja Kroke, and Heiner Boeing.
\newblock An approach to construct simplified measures of dietary patterns from exploratory factor analysis.
\newblock \emph{British Journal of Nutrition}, 89\penalty0 (3):\penalty0 409--418, 2003.

\bibitem[Edefonti et~al.(2012)]{edefonti2012nutrient}
Valeria Edefonti et~al.
\newblock Nutrient-based dietary patterns and the risk of head and neck cancer: a pooled analysis in the international head and neck cancer epidemiology consortium.
\newblock \emph{Annals of Oncology}, 23\penalty0 (7):\penalty0 1869--1880, 2012.

\bibitem[Bennett et~al.(2022)Bennett, Bardon, and Gibney]{bennett2022comparison}
Grace Bennett, Laura~A Bardon, and Eileen~R Gibney.
\newblock A comparison of dietary patterns and factors influencing food choice among ethnic groups living in one locality: a systematic review.
\newblock \emph{Nutrients}, 14\penalty0 (5):\penalty0 941, 2022.

\bibitem[Castello et~al.(2014)]{castello2014spanish}
Adela Castello et~al.
\newblock Spanish mediterranean diet and other dietary patterns and breast cancer risk: case--control epigeicam study.
\newblock \emph{British journal of cancer}, 111\penalty0 (7):\penalty0 1454--1462, 2014.

\bibitem[Castell{\'o} et~al.(2022)]{castello2022adherence}
Adela Castell{\'o} et~al.
\newblock Adherence to the western, prudent and mediterranean dietary patterns and colorectal cancer risk: Findings from the spanish cohort of the european prospective investigation into cancer and nutrition (epic-spain).
\newblock \emph{Nutrients}, 14\penalty0 (15):\penalty0 3085, 2022.

\bibitem[De~Vito et~al.(2019{\natexlab{a}})De~Vito, Bellio, Trippa, and Parmigiani]{DeVito2019Mfa}
Roberta De~Vito, Ruggero Bellio, Lorenzo Trippa, and Giovanni Parmigiani.
\newblock Multi-study factor analysis.
\newblock \emph{Biometrics}, 75\penalty0 (1):\penalty0 337--346, 2019{\natexlab{a}}.
\newblock ISSN 0006-341X.

\bibitem[Roy et~al.(2021)Roy, Lavine, Herring, and Dunson]{roy2021perturbed}
Arkaprava Roy, Isaac Lavine, Amy~H Herring, and David~B Dunson.
\newblock Perturbed factor analysis: Accounting for group differences in exposure profiles.
\newblock \emph{The annals of applied statistics}, 15\penalty0 (3):\penalty0 1386, 2021.

\bibitem[Chandra et~al.(2024)Chandra, Dunson, and Xu]{chandra2024inferring}
Noirrit~Kiran Chandra, David~B Dunson, and Jason Xu.
\newblock Inferring covariance structure from multiple data sources via subspace factor analysis.
\newblock \emph{Journal of the American Statistical Association}, pages 1--15, 2024.

\bibitem[De~Vito et~al.(2019{\natexlab{b}})]{de2019shared}
Rroberta De~Vito et~al.
\newblock Shared and study-specific dietary patterns and head and neck cancer risk in an international consortium.
\newblock \emph{Epidemiology (Cambridge, Mass.)}, 30\penalty0 (1):\penalty0 93, 2019{\natexlab{b}}.

\bibitem[Hashibe et~al.(2007)]{hashibe2007alcohol}
Mia Hashibe et~al.
\newblock Alcohol drinking in never users of tobacco, cigarette smoking in never drinkers, and the risk of head and neck cancer: pooled analysis in the international head and neck cancer epidemiology consortium.
\newblock \emph{Journal of the National Cancer Institute}, 99\penalty0 (10):\penalty0 777--789, 2007.

\bibitem[Conway et~al.(2009)]{conway2009enhancing}
David~I Conway et~al.
\newblock Enhancing epidemiologic research on head and neck cancer: Inhance-the international head and neck cancer epidemiology consortium.
\newblock \emph{Oral oncology}, 45\penalty0 (9):\penalty0 743--746, 2009.

\bibitem[De~Vito et~al.(2022)]{de2022shared}
Roberta De~Vito et~al.
\newblock Shared and ethnic background site-specific dietary patterns in the hispanic community health study/study of latinos (hchs/sol).
\newblock \emph{medRxiv}, pages 2022--06, 2022.

\bibitem[Sorlie et~al.(2010)]{sorlie2010design}
Paul~D Sorlie et~al.
\newblock Design and implementation of the hispanic community health study/study of latinos.
\newblock \emph{Annals of epidemiology}, 20\penalty0 (8):\penalty0 629--641, 2010.

\bibitem[Stephenson et~al.(2020)Stephenson, Herring, and Olshan]{stephenson2020robust}
Briana~JK Stephenson, Amy~H Herring, and Andrew Olshan.
\newblock Robust clustering with subpopulation-specific deviations.
\newblock \emph{Journal of the American Statistical Association}, 115\penalty0 (530):\penalty0 521--537, 2020.

\bibitem[Yoon et~al.(2001)]{yoon2001national}
Paula~W Yoon et~al.
\newblock The national birth defects prevention study.
\newblock \emph{Public health reports}, 116\penalty0 (Suppl 1):\penalty0 32, 2001.

\bibitem[Stephenson et~al.(2022)Stephenson, Herring, and Olshan]{stephenson2022derivation}
Briana~JK Stephenson, Amy~H Herring, and Andrew~F Olshan.
\newblock Derivation of maternal dietary patterns accounting for regional heterogeneity.
\newblock \emph{Journal of the Royal Statistical Society Series C: Applied Statistics}, 71\penalty0 (5):\penalty0 1957--1977, 2022.

\bibitem[Trichopoulos et~al.(1985)]{trichopoulos1985diet}
D~Trichopoulos et~al.
\newblock Diet and cancer of the stomach: a case-control study in {Greece}.
\newblock \emph{International journal of cancer}, 36\penalty0 (3):\penalty0 291--297, 1985.

\bibitem[Tucker et~al.(1997)Tucker, Seljaas, and Hager]{tucker1997body}
Larry~A Tucker, Gary~T Seljaas, and Ronald~L Hager.
\newblock Body fat percentage of children varies according to their diet composition.
\newblock \emph{Journal of the American Dietetic Association}, 97\penalty0 (9):\penalty0 981--986, 1997.

\bibitem[West(2003)]{2003Bfrm}
Mike West.
\newblock \emph{Bayesian factor regression models in the ``large p, small n'' paradigm}, volume~7.
\newblock Elsevier, North-Holland [u.a.], 2003.
\newblock ISBN 0444877460.

\bibitem[Carvalho et~al.(2008)]{Carvalho2008HSFM}
Carlos~M. Carvalho et~al.
\newblock High-dimensional sparse factor modeling: Applications in gene expression genomics.
\newblock \emph{Journal of the American Statistical Association}, 103\penalty0 (484):\penalty0 1438--1456, 2008.
\newblock ISSN 0162-1459.

\bibitem[Avalos-Pacheco et~al.(2022)Avalos-Pacheco, Rossell, and Savage]{Avalos2022HLDI}
Alejandra Avalos-Pacheco, David Rossell, and Richard~S. Savage.
\newblock {Heterogeneous large datasets integration using {Bayesian} factor regression}.
\newblock \emph{Bayesian analysis}, 17\penalty0 (1):\penalty0 33--66, 2022.
\newblock ISSN 1936-0975.

\bibitem[{National Heart, Lung, and Blood Institute and others}(2009)]{national2009hispanic}
{National Heart, Lung, and Blood Institute and others}.
\newblock Hispanic community health study/study of latinos (hchs/sol).
\newblock \emph{Bethesda, MD: NHLBI}, 2009.

\bibitem[Daviglus et~al.(2014)Daviglus, Pirzada, and Talavera]{daviglus2014cardiovascular}
Martha~L Daviglus, Amber Pirzada, and Gregory~A Talavera.
\newblock Cardiovascular disease risk factors in the hispanic/latino population: lessons from the hispanic community health study/study of latinos (hchs/sol).
\newblock \emph{Progress in cardiovascular diseases}, 57\penalty0 (3):\penalty0 230--236, 2014.

\bibitem[Sofianou et~al.(2011)Sofianou, Fung, and Tucker]{sofianou2011differences}
Anastasia Sofianou, Teresa~T Fung, and Katherine~L Tucker.
\newblock Differences in diet pattern adherence by nativity and duration of us residence in the mexican-american population.
\newblock \emph{Journal of the American Dietetic Association}, 111\penalty0 (10):\penalty0 1563--1569, 2011.

\bibitem[Batis et~al.(2011)Batis, Hernandez-Barrera, Barquera, Rivera, and Popkin]{batis2011food}
Carolina Batis, Lucia Hernandez-Barrera, Simon Barquera, Juan~A Rivera, and Barry~M Popkin.
\newblock Food acculturation drives dietary differences among mexicans, mexican americans, and non-hispanic whites.
\newblock \emph{The Journal of nutrition}, 141\penalty0 (10):\penalty0 1898--1906, 2011.

\bibitem[Davis et~al.(2013)Davis, Schechter, Ortega, Rosen, Wylie-Rosett, and Walker]{davis2013dietary}
Nichola~J Davis, Clyde~B Schechter, Felix Ortega, Rosa Rosen, Judith Wylie-Rosett, and Elizabeth~A Walker.
\newblock Dietary patterns in blacks and hispanics with diagnosed diabetes in new york city’s south bronx.
\newblock \emph{The American journal of clinical nutrition}, 97\penalty0 (4):\penalty0 878--885, 2013.

\bibitem[Maldonado et~al.(2022)Maldonado, Sotres-Alvarez, Mattei, Daviglus, Talavera, Perreira, Van~Horn, Mossavar-Rahmani, LeCroy, Gallo, et~al.]{maldonado2022posteriori}
Luis~E Maldonado, Daniela Sotres-Alvarez, Josiemer Mattei, Martha~L Daviglus, Gregory~A Talavera, Krista~M Perreira, Linda Van~Horn, Yasmin Mossavar-Rahmani, Madison~N LeCroy, Linda~C Gallo, et~al.
\newblock A posteriori dietary patterns, insulin resistance, and diabetes risk by hispanic/latino heritage in the hchs/sol cohort.
\newblock \emph{Nutrition \& Diabetes}, 12\penalty0 (1):\penalty0 44, 2022.

\bibitem[Matheus et~al.(2013)Matheus, Tannus, Cobas, Palma, Negrato, and Gomes]{matheus2013impact}
Alessandra Saldanha de~Mattos Matheus, Lucianne Righeti~Monteiro Tannus, Roberta~Arnoldi Cobas, Catia C~Sousa Palma, Carlos~Antonio Negrato, and Marilia de~Brito Gomes.
\newblock Impact of diabetes on cardiovascular disease: an update.
\newblock \emph{International journal of hypertension}, 2013\penalty0 (1):\penalty0 653789, 2013.

\bibitem[Berger et~al.(2015)Berger, Raman, Vishwanathan, Jacques, and Johnson]{berger2015dietary}
Samantha Berger, Gowri Raman, Rohini Vishwanathan, Paul~F Jacques, and Elizabeth~J Johnson.
\newblock Dietary cholesterol and cardiovascular disease: a systematic review and meta-analysis.
\newblock \emph{The American journal of clinical nutrition}, 102\penalty0 (2):\penalty0 276--294, 2015.

\bibitem[Sowers et~al.(2001)Sowers, Epstein, and Frohlich]{sowers2001diabetes}
James~R Sowers, Murray Epstein, and Edward~D Frohlich.
\newblock Diabetes, hypertension, and cardiovascular disease: an update.
\newblock \emph{Hypertension}, 37\penalty0 (4):\penalty0 1053--1059, 2001.

\bibitem[Liese et~al.(2022)]{liese2022variations}
Angela~D Liese et~al.
\newblock Variations in dietary patterns defined by the healthy eating index 2015 and associations with mortality: findings from the dietary patterns methods project.
\newblock \emph{The Journal of Nutrition}, 152\penalty0 (3):\penalty0 796--804, 2022.

\bibitem[De~Vito et~al.(2021)De~Vito, Bellio, Trippa, and Parmigiani]{DeVito2021Bmfa}
Roberta De~Vito, Ruggero Bellio, Lorenzo Trippa, and Giovanni Parmigiani.
\newblock Bayesian multistudy factor analysis for high-throughput biological data.
\newblock \emph{The annals of applied statistics}, 15\penalty0 (4):\penalty0 1723--1741, 2021.
\newblock ISSN 1932-6157.

\bibitem[Avalos~Pacheco(2018)]{Avalos2018Frfd}
Alejandra Avalos~Pacheco.
\newblock \emph{Factor regression for dimensionality reduction and data integration techniques with applications to cancer data}.
\newblock University of {Warwick}, {PhD} thesis, 2018.

\bibitem[Meng and Rubin(1993)]{MengECM93}
Xiao-Li Meng and Donald~B. Rubin.
\newblock Maximum likelihood estimation via the ecm algorithm: A general framework.
\newblock \emph{Biometrika}, 80\penalty0 (2):\penalty0 267--278, 1993.
\newblock ISSN 0006-3444.

\bibitem[Morgenstern and Woodbury(1950)]{Morgenstern1950SoIo}
Oskar Morgenstern and Max~A Woodbury.
\newblock Stability of inverses of input-output matrices.
\newblock \emph{Econometrica}, 18:\penalty0 190, 1950.
\newblock ISSN 0012-9682.

\bibitem[McLachlan(2008)]{McLachlan2008TEaa}
Geoffrey~J. McLachlan.
\newblock \emph{The EM algorithm and extensions}.
\newblock Wiley series in probability and statistics. Wiley-Interscience, Hoboken, N.J., 2nd ed. edition, 2008.
\newblock ISBN 9780471201700.

\bibitem[Akaike(1974)]{AkaikeH1974Anla}
Hirotsugu Akaike.
\newblock A new look at the statistical model identification.
\newblock \emph{IEEE transactions on automatic control}, 19\penalty0 (6):\penalty0 716--723, 1974.
\newblock ISSN 0018-9286.

\bibitem[Schwarz(1978)]{SCHWARZG1978EtDo}
Gideon Schwarz.
\newblock Estimating the dimension of a model.
\newblock \emph{The Annals of statistics}, 6\penalty0 (2):\penalty0 461--464, 1978.
\newblock ISSN 0090-5364.

\bibitem[Kaiser(1958)]{Kaiser1958Tvcf}
Henry~F. Kaiser.
\newblock The varimax criterion for analytic rotation in factor analysis.
\newblock \emph{Psychometrika}, 23\penalty0 (3):\penalty0 187--200, 1958.
\newblock ISSN 0033-3123.

\bibitem[Lopes and West(2004)]{Lopes2004BMAI}
Hedibert~Freitas Lopes and Mike West.
\newblock Bayesian model assessment in factor analysis.
\newblock \emph{Statistica Sinica}, 14\penalty0 (1):\penalty0 41--67, 2004.
\newblock ISSN 1017-0405.

\bibitem[Fr\"uhwirth-Schnatter et~al.(2023)Fr\"uhwirth-Schnatter, Hosszejni, and Lopes]{FruehwirthSchnatter2023SBfa}
Sylvia Fr\"uhwirth-Schnatter, Darjus Hosszejni, and Hedibert~Freitas Lopes.
\newblock Sparse bayesian factor analysis when the number of factors is unknown.
\newblock \emph{arXiv}, 2023.

\bibitem[Robert(1976)]{robert1976}
P.~Robert.
\newblock A unifying tool for linear multivariate statistical methods : The rv-coefficient.
\newblock \emph{J. R. Stat. Soc. Ser. C}, 25:\penalty0 257--265, 1976.

\bibitem[Kjeldsen(2018)]{kjeldsen2018hypertension}
Sverre~E Kjeldsen.
\newblock Hypertension and cardiovascular risk: General aspects.
\newblock \emph{Pharmacological research}, 129:\penalty0 95--99, 2018.

\bibitem[DiStefano et~al.(2009)DiStefano, Zhu, and Mindrila]{distefano2009understanding}
Christine DiStefano, Min Zhu, and Diana Mindrila.
\newblock Understanding and using factor scores: Considerations for the applied researcher.
\newblock \emph{Practical Assessment, Research, and Evaluation}, 14\penalty0 (1):\penalty0 20, 2009.

\bibitem[Lumley and Scott(2017)]{lumley2017fitting}
Thomas Lumley and Alastair Scott.
\newblock Fitting regression models to survey data.
\newblock \emph{Statistical Science}, pages 265--278, 2017.

\bibitem[Davy et~al.(2002)Davy, Davy, Ho, Beske, Davrath, and Melby]{davy2002high}
Brenda~M Davy, Kevin~P Davy, Richard~C Ho, Stacy~D Beske, Linda~R Davrath, and Christopher~L Melby.
\newblock High-fiber oat cereal compared with wheat cereal consumption favorably alters ldl-cholesterol subclass and particle numbers in middle-aged and older men.
\newblock \emph{The American journal of clinical nutrition}, 76\penalty0 (2):\penalty0 351--358, 2002.

\bibitem[Clifton and Keogh(2018)]{clifton2018cholesterol}
Peter Clifton and Jennifer Keogh.
\newblock Cholesterol-lowering effects of plant sterols in one serve of wholegrain wheat breakfast cereal biscuits—a randomised crossover clinical trial.
\newblock \emph{Foods}, 7\penalty0 (3):\penalty0 39, 2018.

\bibitem[Vitale et~al.(2019)Vitale, Masulli, Rivellese, Bonora, Babini, Sartore, Corsi, Buzzetti, Citro, Baldassarre, et~al.]{vitale2019pasta}
Marilena Vitale, Maria Masulli, Angela~Albarosa Rivellese, Enzo Bonora, Anna~Carla Babini, Giovanni Sartore, Laura Corsi, Raffaella Buzzetti, Giuseppe Citro, Maria Pompea~Antonia Baldassarre, et~al.
\newblock Pasta consumption and connected dietary habits: associations with glucose control, adiposity measures, and cardiovascular risk factors in people with type 2 diabetes—tosca. it study.
\newblock \emph{Nutrients}, 12\penalty0 (1):\penalty0 101, 2019.

\bibitem[Glazener et~al.(1964)Glazener, Morgan, Simpson, and Johnson]{glazener1964pargyline}
Frederic~S Glazener, William~A Morgan, John~M Simpson, and Paul~K Johnson.
\newblock Pargyline, cheese, and acute hypertension.
\newblock \emph{JAMA}, 188\penalty0 (8):\penalty0 754--755, 1964.

\bibitem[Hansen et~al.(2024)Hansen, Avalos-Pacheco, Russo, and De~Vito]{hansen2024fast}
Blake Hansen, Alejandra Avalos-Pacheco, Massimiliano Russo, and Roberta De~Vito.
\newblock Fast variational inference for bayesian factor analysis in single and multi-study settings.
\newblock \emph{Journal of Computational and Graphical Statistics}, \penalty0 (just-accepted):\penalty0 1--42, 2024.

\bibitem[Fruhwirth-Schnatter et~al.(2024)Fruhwirth-Schnatter, Hosszejni, and Lopes]{Schnatter}
S.~Fruhwirth-Schnatter, D.~Hosszejni, and H.~F. Lopes.
\newblock Sparse bayesian factor analysis when the number of factors is unknown.
\newblock \emph{Bayesian Analysis}, 1\penalty0 (1):\penalty0 1--31, 2024.

\bibitem[Tyekucheva et~al.(2011)Tyekucheva, Marchionni, Karchin, and Parmigiani]{tyekucheva2011integrating}
Svitlana Tyekucheva, Luigi Marchionni, Rachel Karchin, and Giovanni Parmigiani.
\newblock Integrating diverse genomic data using gene sets.
\newblock \emph{Genome biology}, 12:\penalty0 1--14, 2011.

\end{thebibliography}

\end{document}